\documentclass[11pt]{article}
\usepackage{graphicx, amssymb}
\title{Spectral Estimates and Non-Selfadjoint Perturbations of
Spheroidal Wave Operators}
\author{Felix Finster and Harald Schmid}
\date{May 2004 / May 2005}

\newtheorem{Def}{Def.}[section]
\newtheorem{Thm}[Def]{Theorem}

\newtheorem{Lemma}[Def]{Lemma}

\newcommand{\Proof}{{\em{Proof. }}}
\newcommand{\QED}{\ \hfill $\FBox$ \\[1em]}
\newcommand{\spc}{\;\;\;\;\;\;\;\;\;\;}

\newcommand{\C}{\mathbb{C}}
\newcommand{\R}{\mathbb{R}}
\newcommand{\1}{\mbox{\rm 1 \hspace{-1.05 em} 1}}
\newcommand{\Z}{\mathbb{Z}}

\newcommand{\N}{\mathbb{N}}

\newcommand{\beq}{\begin{equation}}
\newcommand{\eeq}{\end{equation}}

\newcommand{\umax}{u_{\mbox{\tiny{max}}}}
\newcommand{\Vmax}{V_{\mbox{\tiny{max}}}}

\newcommand{\FBox}{\rule{2mm}{2.25mm}}

\begin{document}
\maketitle

\begin{abstract}
We derive a spectral representation for the oblate spheroidal wave operator
which is holomorphic in the aspherical parameter $\Omega$ in a neighborhood
of the real line. For real $\Omega$, estimates are derived for all eigenvalue
gaps uniformly in~$\Omega$.

The proof of the gap estimates is based on detailed estimates for complex
solutions of the Riccati equation. The spectral representation for complex
$\Omega$ is obtained using the theory of slightly non-selfadjoint perturbations.
\end{abstract}

\section{Introduction}
\setcounter{equation}{0} \label{sec1}
Recently an integral representation was derived for solutions of the scalar
wave equation in the Kerr black hole geometry~\cite{FKSY}. This result
relies crucially on a spectral representation for the oblate spheroidal wave
operator for complex values of the aspherical parameter~$\Omega$ (also
referred to as ``ellipticity parameter'' or ``semifocal distance'').
In the present paper, this spectral representation is proved. The reason
why this problem deserves to be worked out in a
separate paper is that most of our methods apply in a much more general context.
Namely, the core of the paper is to derive estimates for the eigenvalue
gaps~$\lambda_{n+1}-\lambda_n$ for real~$\Omega$, which are uniform in~$\Omega$
and~$n$. To this end, we need to control the eigenvalues and the behavior of
the wave functions in detail. Our method is based on invariant region estimates
for the complex Riccati equation and applies to general
Sturm-Liouville or one-dimensional Schr{\"o}dinger problems. In particular, it gives refined error estimates for
WKB approximations. We regard the spheroidal wave equation as a model
problem for working out these estimates.

Despite the vast literature on spectral estimates for the Schr{\"o}dinger equation
(see e.g.~\cite{RS} and the references therein), gap estimates are rarely
found in the standard literature. Most papers are concerned with the two lowest
eigenvalues~\cite{Lavine, Y}, or they apply in special situations like for a
a nearly constant potential~\cite{LH}. Probably, this is because gap
estimates depend sensitively on the detailed form of the potential (as one
sees in the example of a double-well potential), making it difficult to get
general results. Our method requires that the potential is piecewise monotone
and that we have good control of its derivatives.

We now introduce our problem and state our results. The spheroidal wave equation
is the eigenvalue equation for the spheroidal wave operator,
a linear elliptic operator with smooth
coefficients on the unit sphere~$S^2$.
Since the spheroidal wave operator is axisymmetric, we can choose angular
variables~$\vartheta \in (0, \pi)$ and~$\varphi \in [0, 2 \pi)$ (with~$\vartheta$ the angle to the axis of symmetry)
and separate out the~$\varphi$-dependence with the plane wave ansatz
$\phi(\vartheta, \varphi) = e^{i k \varphi}\:\Theta(\vartheta)$, $k \in \Z$. After this separation, the
spheroidal wave equation takes the form
\begin{equation} \label{swe}
{\cal{A}}\,\Theta \;=\; \lambda \, \Theta\:,
\end{equation}
where ${\cal{A}}$ is the linear differential operator of second order
\begin{equation} \label{swo}
{\cal{A}} \;=\;
-\frac{d}{d \cos \vartheta}\: \sin^2 \vartheta\:
\frac{d}{d \cos \vartheta}
+\frac{1}{\sin^2 \vartheta}(\Omega \sin^2 \vartheta + k )^{2}
\end{equation}
on the interval~$\vartheta \in (0, \pi)$. Here~$\Omega \in \C$ is the
aspherical parameter.
In the special case~$\Omega=0$, the spheroidal wave operator
simplifies to the spherical Laplacian, and the Legendre
polynomials~$P^k_l(\cos \vartheta)$ are explicit solutions to~(\ref{swe}).
We shall consider the spheroidal wave equation for fixed~$k$, but for
a variable complex parameter~$\Omega$.
The fact that the eigenfunction~$\phi$ should be smooth at the
poles~$\vartheta =0, \pi$ of the sphere gives rise to the following
boundary conditions,
\beq \label{bc}
\left\{ \begin{array}{rclcl}
\displaystyle \lim_{\vartheta \to 0, \pi}
\Theta'(\vartheta) &=& 0 &\qquad& {\mbox{if $k=0$}} \\[.5em]
\displaystyle \lim_{\vartheta \to 0, \pi}
\Theta(\vartheta) &=& 0 &\;\;\;& {\mbox{if $k \neq 0$}}\:. \end{array}
\right.
\eeq
We consider~${\cal{A}}$ as an operator in the Hilbert space
${\cal{H}}=L^2((0, \pi), \sin \vartheta \,d\vartheta)$ with
domain of definition given by those functions
in $C^2(0, \pi)$ which satisfy the boundary conditions~(\ref{bc}).
Note that the potential in the spheroidal wave operator
is in general complex,
\begin{equation} \label{imaginary}
{\mbox{Im}} \left( \frac{(\Omega \sin^2 \vartheta + k )^{2}}{\sin^2 \vartheta}
\right) \;=\; 2 \left( {\mbox{Re}}\, \Omega\:
\sin^2 \vartheta + k \right) {\mbox{Im}}\, \Omega \:,
\end{equation}
and therefore~${\cal{A}}$ is symmetric only if~$\Omega$ is real.
In previous works, asymptotic expansions for individual eigenvalues are
derived~\cite{Flammer, MS}, and it is shown numerically that eigenvalues
can degenerate for non-real~$\Omega$~\cite{HG}, but rigorous estimates or
completeness statements are not given.
Our main result is the following spectral representation for~$\Omega$
in a neighborhood of the real line.
\begin{Thm} \label{thm1}
For any $k \in \Z$ and $c>0$, we define the open set $U \subset \C$ by the condition
\beq \label{ocond}
|{\mbox{\rm{Im}}}\, \Omega| \;<\; \frac{c}{1 + |{\mbox{\rm{Re}}}\, \Omega|}\:.
\eeq
Then there is a positive integer~$N$ and a family of
operators~$Q_n(\Omega)$ on~${\cal{H}}$
defined for $n \in \N \cup \{0\}$ and $\Omega \in U$
with the following properties:
\begin{description}
\item[(i)]
The~$Q_n$ are holomorphic in~$\Omega$.
\item[(ii)]
$Q_0$ is a projector on an $N$-dimensional invariant
subspace of ${\cal{A}}$. For $n>0$, the $Q_n$ are projectors
on one-dimensional eigenspaces of ${\cal{A}}$ with corresponding
eigenvalues $\lambda_n(\Omega)$. These eigenvalues satisfy a bound of
the form
\beq \label{ange1}
|\lambda_n(\Omega)| \;\leq\; C(n)\: (1+|\Omega|)
\eeq
for suitable constants $C(n)$. Furthermore, there is a parameter
$\varepsilon>0$ such that for all $n \in \N$ and $\Omega \in U$,
\beq \label{ang11}
|\lambda_n(\Omega)| \;\geq\; n\: \varepsilon\:.
\eeq
\item[(iii)] The $Q_n$ are complete, i.e.\
\[ \sum_{k=0}^\infty Q_n \;=\; \1 \]
with strong convergence of the series.
\item[(iv)] The $Q_n$ are uniformly bounded,
i.e. for all $n \in \N \cup \{0\}$,
\beq \label{Qnb}
\|Q_n\| \;\leq\; c_1
\eeq
with $c_1$ independent of $\Omega$ and $k$.
\end{description}
If $c$ is sufficiently small, $c<\delta$, or
the real part of $\Omega$ is sufficiently large,
$|{\mbox{\em{Re}}}\, \Omega| > C(c)$, one can choose $N=1$, i.e.\
${\cal{A}}$ has a purely discrete spectrum consisting of simple eigenvalues.
\end{Thm}
To avoid misunderstandings, we point out that by a ``projector on an invariant
subspace of~${\cal{A}}$'' we mean an operator~$Q$ which is idempotent and
commutes with~${\cal{A}}$. But~$Q$ will in general not be symmetric.

In our proof we shall treat the imaginary part of the potential~(\ref{imaginary})
as a slightly non-selfadjoint perturbation in the spirit
of~\cite[V.4.5]{Kato}, see also~\cite[Chapter~12]{CL}.
For this method to be applicable, we need good control of the eigenvalues
of the corresponding selfadjoint problem. Our starting point is the following
spectral decomposition of ${\cal{A}}$ in the case of {\em{real}}~$\Omega$.
\begin{Thm} \label{thm2}
For any~$k \in \Z$ and~$\Omega \in \R$, the operator~${\cal{A}}$
has a unique selfadjoint extension compatible with the boundary
conditions~(\ref{bc}). This extension, which we again denote by~${\mathcal{A}}$,
is a positive operator with compact resolvent and simple eigenvalues.
It is invariant on the even and odd parity subspaces ${\cal{H}}^\pm$
defined by
\[ {\cal{H}} \;=\; {\cal{H}}^+ \oplus {\cal{H}}^-
\spc {\mbox{with}} \spc {\cal{H}}^\pm = \{ \phi \in {\cal{H}}
{\mbox{ with }} \phi(\pi-\vartheta) = \pm\phi(\vartheta) \}. \]
\end{Thm}
We denote the eigenvalues of~${\cal{A}}$ restricted to~${\cal{H}}^\pm$
by~$\lambda^\pm_n$ and count them with multiplicities,
\[ 0 \;\leq\; \lambda^\pm_1 \;<\; \lambda^\pm_2
\;<\; \lambda^\pm_3 \;<\; \cdots . \]

Using abstract methods (see \cite[Theorem~3.9, VII.3.5]{Kato}),
one could show that each eigenvalue~$\lambda^\pm_n(\Omega)$ has a holomorphic continuation
to a neighborhood of the real axis. However, as pointed out
in~\cite[Remark~3.9, VII.3.5]{Kato}, this neighborhood will
depend on~$n$, making it impossible construct a
neighborhood in which all the~$\lambda^\pm_n(\Omega)$ exist.
Therefore, abstract methods only seem to give results which are much weaker
than Theorem~\ref{thm1}, where the whole spectral decomposition is
shown to have a holomorphic continuation to a neighborhood of the
real axis.
Furthermore, we point out that the parameter~$c$ in the statement of
Theorem~\ref{thm1} can be chosen arbitrarily large.
Therefore, the holomorphic family of operators~$Q_n(\Omega)$ is not only
defined in a small neighborhood of the real axis, but in
a strip~(\ref{ocond}) which can enclose any bounded subset of the
complex plane. The key for getting this strong result are
the following gap estimates uniform in~$n$ and~$\Omega$.
\begin{Thm} \label{thm3}
For any~$k \in \Z$ and~$\gamma>0$, there is a positive integer~$N$ such that
\[ \lambda^\pm_{n+1} - \lambda^\pm_n \;>\; \gamma
\spc {\mbox{for all $n \geq N$ and $\Omega \in \R$}}. \]
If $\gamma$ is sufficiently small or
$|\Omega|$ is sufficiently large, one can choose $N=1$.
\end{Thm}

The paper is organized as follows. In Section~\ref{sec2} we prove
Theorem~\ref{thm2} and reduce Theorem~\ref{thm3} to gap estimates
for a self-adjoint Sturm-Liouville operator on the interval~$\vartheta \in
(0, \frac{\pi}{2}]$ with suitable boundary conditions.
In Sections~\ref{sec2}--\ref{sec6} we introduce the complex
Riccati equation and develop general techniques for analyzing
its solutions. In Section~\ref{sec7} and~\ref{sec8} we apply these
techniques to the spheroidal wave operator and prove Theorem~\ref{thm3}.
Finally, in Section~\ref{sec8} we use perturbative methods to
proof Theorem~\ref{thm1}.

\section{Basic Considerations}
\setcounter{equation}{0}
\label{sec2}
\setcounter{equation}{0}
Until the end of Section~\ref{sec7} we will
consider the spheroidal wave equation~(\ref{swe})
for real~$\Omega$. Using that~(\ref{swo}) is invariant under the
transformations $\Omega \to -\Omega$ and $k \to -k$,
we can assume throughout that
\[ \Omega \;>\: 0 \:. \]

Let us derive a spectral representation of the spheroidal wave operator.
One possible method would be to apply elliptic theory to the spheroidal
wave operator on~$S^2$ before separation of variables.
After choosing a self-adjoint extension on the Hilbert space~$L^2(S^2)$,
one could apply the abstract spectral theorem, and projecting
the resulting smooth eigenfunctions on the subspace for fixed~$k$
would give the desired spectral decomposition for the ordinary differential operator~(\ref{swo}).
For clarity, we will in this paper restrict attention to ODE techniques.
Thus we avoid elliptic theory and prefer to apply Sturm-Liouville theory.
In the variable $u=\vartheta \in (0, \pi)$, the operator~(\ref{swo})
can be written as
\[ {\cal{A}} \;=\; -\frac{1}{\sin u}\: \frac{d}{d u}\: \sin u\:
\frac{d}{d u}
+\frac{1}{\sin^2 u}(\Omega \sin^2 u + k )^{2}. \]
In order to bring this operator to the standard
Sturm-Liouville form, we introduce the function $Y$ by
\beq \label{Ydef}
Y \;=\; \sqrt{\sin u}\: \Theta \:.
\eeq
Then
\[ B\: Y \;=\; \lambda \:Y \;, \]
where
\begin{eqnarray*}
B &=& -\frac{1}{\sqrt{\sin u}}\: \frac{d}{d u}\: \sin u\:
\frac{d}{d u}\:\frac{1}{\sqrt{\sin u}}
+\frac{1}{\sin^2 u}(\Omega \sin^2 u + k )^{2} \\
&=& -\frac{d^2}{d u^2} + \frac{1}{2}\: \frac{\cos^2 u}{\sin^2 u}
-\sqrt{\sin u} \: \left(\frac{1}{\sqrt{\sin u}} \right)''
+\frac{1}{\sin^2 u}(\Omega \sin^2 u + k )^{2} \\
&=& -\frac{d^2}{d u^2} - \frac{1}{4}\: \frac{\cos^2 u}{\sin^2 u}
- \frac{1}{2} +\frac{1}{\sin^2 u}(\Omega \sin^2 u + k )^{2} \:.
\end{eqnarray*}
Thus $Y$ satisfies the Sturm-Liouville equation
\begin{equation} \label{5ode}
\left( -\frac{d^2}{du^2} + V \right) Y \;=\; 0 \:,
\end{equation}
where~$V$ is the potential
\begin{equation} \label{Vdef}
V \;=\; \Omega^2\: \sin^2 u + \left(k^2-\frac{1}{4} \right) \frac{1}{\sin^2 u}
\:-\: \mu
\end{equation}
with~$\mu$ the constant
\begin{equation}
\mu \;=\; \lambda - 2 \Omega k + \frac{1}{4} . \label{mudef}
\end{equation}
The transformation~(\ref{Ydef}) from~$\Theta$ to~$Y$ becomes
a unitary transformation if the integration measure in the
corresponding Hilbert spaces is transformed from~$\sin u\:du$ to~$du$.
Thus the eigenvalue problem~(\ref{swe}) on~${\mathcal{H}}$ is equivalent
to~(\ref{5ode}) on the Hilbert space~$L^2((0,\pi), du)$.
The boundary conditions~(\ref{bc}) at~$u=0$ (and similarly at~$u=\pi$)
can be written as
\beq \label{bcb}
\left\{ \begin{array}{rclcl}
\displaystyle \lim_{u \searrow 0}
(u^{-\frac{1}{2}}\: Y)'(u) &=& 0 &\qquad& {\mbox{if $k=0$}} \\[.5em]
\displaystyle \lim_{u \searrow 0}
u^{-\frac{1}{2}}\: Y(u) &=& 0 && {\mbox{if $k \neq 0$}}\:.
\end{array} \right.
\eeq

The Sturm-Liouville equation~(\ref{5ode}) is singular at
the two end points~$u=0, \pi$. An asymptotic expansion near~$u=0$ shows that~(\ref{5ode})
has fundamental solutions~$Y_{1\!/\!2}$ of the following form,
\beq \label{asymp}
\left\{ \begin{array}{clccll}
Y_1 = \sqrt{u} + {\mathcal{O}}(u^{\frac{3}{2}}) &,\quad&
Y_2 = \sqrt{u}\: \log(u) + {\mathcal{O}}(u^{\frac{1}{2}})
 &\qquad& {\mbox{if $k=0$}} \\[.5em]
Y_1 = u^{\frac{1}{2}+|k|} + {\mathcal{O}}(u^{\frac{3}{2}+|k|}) &,&
Y_2 = u^{\frac{1}{2}-|k|} + {\mathcal{O}}(u^{\frac{3}{2}-|k|})
&& {\mbox{if $k \neq 0$}}\:,
\end{array} \right.
\eeq
and similary at~$u=\pi$.
In the case~$k \neq 0$, $Y_1$ is square integrable near~$u=0$,
whereas~$Y_2$ is not. Thus, using Weyl's notation, the Sturm-Liouville
operator is in the limit point case at both end points, and
thus~${\mathcal{A}}$ is essentially selfadjoint
(see~\cite[Sections~9.2, 9.3]{CL} or~\cite[Chapter XIII.2]{DS}).
In the case~$k=0$, on the other hand, both fundamental solutions are
square integrable. This is the limit circle case, and the
von-Neumann boundary conditions~(\ref{bcb}) choose a unique
self-adjoint extension (see~\cite[Sections~9.4]{CL}
or~\cite[Chapter XIII.2]{DS}).
We conclude that the Sturm-Liouville operator in~(\ref{5ode})
has a unique self-adjoint extension in~$L^2((0, \pi))$ which
satisfies the boundary conditions~(\ref{bcb}).
Hence the spectral theorem for unbounded operators in Hilbert spaces
gives us the desired spectral representation of~${\mathcal{A}}$.

For each~$\lambda \in \R$, there are (up to a constant) unique
solutions of the ODE which satisfy the boundary conditions at
$u=0$ and $u=\pi$, respectively. If the Wronskian of these
two solutions vanishes, we obtain an eigenfunction in $L^2((0, \pi))$.
Otherwise, these two solutions can be used to define the resolvent
(see~\cite[XIII.3]{DS}), which is compact (see~\cite[XIII.4]{DS}).
This shows that the operator~${\mathcal{A}}$
has a purely discrete spectrum consisting of simple eigenvalues
without limit points. The positivity of~$A$ is
obvious from~(\ref{swo}).

Finally, the boundary value problem~(\ref{5ode}, \ref{bcb}) is
invariant under the transformation~$u \to \pi-u$.
Hence the parity subspaces $L^\pm := \{ \phi \in L^2((0, \pi)) \:|\:
\phi(\pi-u) = \pm \phi(u) \}$ are invariant
under~${\cal{A}}$. This concludes the proof of Theorem~\ref{thm2}.

Clearly, the eigenfunctions~$Y^\pm$ of even and odd parity satisfy
at~$u=\frac{\pi}{2}$ the boundary conditions
\begin{equation} \label{bc2}
\left\{ \begin{array}{c}
\displaystyle (Y^+_n)'\!\left(\frac{\pi}{2}\right)=0 \\[.9em]
\displaystyle Y^-_n\!\left(\frac{\pi}{2}\right)=0 \:. \end{array} \right.
\end{equation}
This makes it possible to consider instead of the the interval~$(0, \pi)$
only the interval~$(0, \frac{\pi}{2}]$ together with the boundary
conditions~(\ref{bcb}, \ref{bc2}).
In what follows, we shall always consider the boundary value
problem~(\ref{5ode}, \ref{bcb}, \ref{bc2}).

In order to better understand Theorem~\ref{thm3}, it is useful to
consider the limits $n \to \infty$ and $\Omega \to \infty$.
For fixed $\Omega$ and large $n$, Weyl's asymptotics applies and yields
that the eigenvalues of ${\cal{A}}$ behave for large~$n$ like the eigenvalues
of the operator~$-\frac{d^2}{du^2}$,
\[ \lambda^\pm_n \;\sim\; n^2 \spc {\mbox{and}} \spc
\lambda^\pm_{n+1}-\lambda^\pm_n \;\sim\; n\:. \]
Therefore, it is obvious that the statement of Theorem~\ref{thm3} holds
for any fixed~$\Omega$ and sufficiently large~$N=N(\Omega)$. The estimate
\beq \label{les}
|\lambda_n(\Omega) - \lambda_n(\Omega')| \;\leq\;
\|{\cal{A}}(\Omega) - {\cal{A}}(\Omega')\|_\infty \;\leq\;
|\Omega-\Omega'|\: (\Omega+\Omega' + 2|k|)
\eeq
yields that eigenvalues of ${\cal{A}}$ are locally Lipschitz in $\Omega$,
uniformly in $n$. This shows that the constant~$N(\Omega)$ can be chosen
locally uniformly in~$\Omega$.
If conversely we fix~$n$, the $n$th spheroidal
eigenvalue~$\lambda_n$ has for large~$\Omega$ the asymptotic expansion
(see~\cite{Flammer} or~\cite{MS})
\beq \label{asymptotics}
\lambda_n(\Omega) =
\left\{ \begin{array}{rl}
2(n+1)\Omega + O(1) & \mbox{ if $n-k$ is even, } \\[1ex]
    2n\Omega + O(1) & \mbox{ if $n-k$ is odd. }  \end{array} \right.
\eeq
Hence for each~$n$, we can make the eigenvalue gap arbitrarily
large by choosing~$\Omega$ sufficiently large. We conclude that it remains
to show that the eigenvalue gaps are bounded uniformly as both $N$ and $|\Omega|$
become large. This is the hard part of Theorem~\ref{thm3}, and
we state it as a separate Lemma.
\begin{Lemma} \label{lemma1}
For any given~$k \in \Z$ and~$c>0$, there are constants~$N \in \N$ and
$\Omega_0>0$ such that
\[ \lambda^\pm_{n+1} - \lambda^\pm_n \;>\; c
\spc {\mbox{for all $n \geq N$ and $\Omega > \Omega_0$}}. \]
\end{Lemma}
The proof of this lemma requires detailed eigenvalue estimates.
We will complete it in Section~\ref{sec7},
and this will also finish the proof of Theorem~\ref{thm3}.

Finally, the node theorem~\cite[Theorem 14.10]{Weidmann} tells us about
the number of zeros of the spheroidal wave functions. In our setting, the statement of the node theorem can easily be derived as follows.
Using the initial conditions~(\ref{bcb}) together with~(\ref{les}),
we obtain from the Picard-Lindel{\"o}f theorem
that the eigenfunctions~$Y^\pm_n$ corresponding to the
eigenvalue~$\lambda^\pm_n$ depend smoothly on the parameter~$\Omega$.
Furthermore, the asymptotics near the boundaries~(\ref{bcb}, \ref{bc2})
shows that the functions~$Y^\pm_n$ have no zeros in
the intervals~$(0, \varepsilon)$ and~$(\frac{\pi}{2}-\varepsilon,
\frac{\pi}{2})$, for a parameter~$\varepsilon>0$ which depends
continuously on~$\Omega$. Thus, if the number of zeros of
the function~$Y^\pm_n$ changed at some~$\Omega$, there would be
a $u \in (0, \pi)$ with $Y^\pm_n(u)=0=(Y^\pm_n)'(u)$, in
contradiction to the fact that~$Y^\pm_n$ does not vanish identically.
We conclude that the number of zeros of~$Y^\pm_n$
on~$(0, \frac{\pi}{2})$ is independent of
$\Omega$, and therefore it suffices to consider the case~$\Omega=0$, when
the spheroidal wave functions reduce to the Legendre polynomials.
Counting their zeros, we conclude that the function
\begin{equation} \label{bc3}
{\mbox{$Y^\pm_n$ has $n$ zeros on $(0, \frac{\pi}{2} )$}} .
\end{equation}

\section{The Complex Riccati Equation}
\setcounter{equation}{0}
\label{sec3}
Let $Y_1$ and $Y_2$ be two real fundamental solutions of the
Sturm-Liouville equation~(\ref{5ode}) for a general real and smooth potential~$V$.
Then their Wronskian
\[ w \;:=\; Y_1(u)\: Y_2'(u) - Y_1'(u) \: Y_2(u) \]
is a constant; we assume in what follows that $w > 0$. We combine the
two real solutions to the complex function
\[ z \;=\; Y_1 + i Y_2 \;, \]
and denote its polar decomposition by
\begin{equation} \label{5g}
z \;=\; \rho\: e^{i \varphi}
\end{equation}
with real functions $\rho(u) \geq 0$ and $\varphi(u)$.
By linearity, $z$ is a complex solution of the Sturm-Liouville equation
\begin{equation} \label{5csch}
z'' \;=\; V \: z \;.
\end{equation}
Note that $z$ has no zeros because at every $u$ at least one of the
fundamental solutions does not vanish. Thus
the function $y$ defined by
\begin{equation}
y \;=\; \frac{z'}{z} \label{5ydef}
\end{equation}
is smooth. Moreover, $y$ satisfies the complex Riccati equation
\begin{equation}
y' + y^2 \;=\; V \;. \label{5c}
\end{equation}
The fact that the solutions of the complex Riccati equation are
smooth will be helpful for getting estimates.
Conversely, from a solution of the Riccati equation one obtains the
corresponding solution of the Sturm-Liouville equation by integration,
\begin{equation} \label{yint}
\left. \log z \right|_{u}^{v} \;=\; \int_{u}^{v} y\;.
\end{equation}
Using~(\ref{5g}) in~(\ref{5ydef}) gives separate equations for the
amplitude and phase of $z$,
\[ \rho' \;=\; \rho\: {\mbox{Re}}\, y \;,\spc
\varphi' \;=\; {\mbox{Im}}\, y \:, \]
and integration gives
\begin{eqnarray}
\left. \log \rho \right|_{u}^{v} &=& \int_{u}^{v} {\mbox{Re}}\; y
\label{5A} \\
\left. \varphi \right|_{u}^{v} &=& \int_{u}^{v} {\mbox{Im}}\; y\;.
\label{5B}
\end{eqnarray}
Furthermore, the Wronskian yields a simple algebraic relation
between~$\rho$ and~$y$. Namely, $w$ can be expressed by
$w = {\mbox{Im}}\: (\overline{z}\: z' )
= \rho^2\: {\mbox{Im}}\:y$ and thus
\begin{equation}
\rho^2 \;=\; \frac{w}{\mbox{Im}\:y}\;. \label{5C}
\end{equation}
Since~$\rho^2$ and~$w$ are non-negative, we see that
\begin{equation}
{\mbox{Im}}\: y(u) > 0 \spc {\mbox{for all $u$}}. \label{5gg}
\end{equation}

The boundary conditions for the Sturm-Liouville equation can easily be translated
into conditions for~$y$. To this end we write the solutions~$Y^\pm_n$ of the Sturm-Liouville
equation corresponding to the eigenvalues~$\lambda^\pm_n$ as~$Y^\pm_n = {\mbox{Im}} (e^{-i \alpha} z^\pm_n)$
with a suitable
phase factor~$e^{-i \alpha}$. Then a Dirichlet condition can be written as
$\varphi = \alpha \, {\mbox{mod}}\, \pi$. A Neumann boundary condition gives
\begin{eqnarray*}
0 &=& {\mbox{Im}} \left( e^{-i \alpha} y z \right) \;=\;
{\mbox{Re}} \left( e^{-i (\alpha+\frac{\pi}{2})} y z \right) \\
&=& \left[{\mbox{Re}} \,y \; \cos(\varphi - \alpha-\frac{\pi}{2}) -
{\mbox{Im}} \,y \; \sin(\varphi - \alpha-\frac{\pi}{2})\right]\rho
\end{eqnarray*}
and thus
\[ \varphi
\;=\; \alpha + \frac{\pi}{2} + \arctan \!\left( \frac{{\mbox{Re}}\, y}{{\mbox{Im}}\, y}
\right) . \]
According to~(\ref{5B}) and~(\ref{5gg}) the function~$\varphi(u)$ is monotone increasing.
Therefore, the number of zeros of $Y$, (\ref{bc3}) tells us how often $\varphi$ crossed
the points~$\, {\mbox{mod}}\, \pi$. This allows us to completely determine the ``phase shifts''
on the interval~$(0, \frac{\pi}{2})$,
\begin{eqnarray}
\varphi^+_n \Big|_0^{\frac{\pi}{2}} &=& \frac{\pi}{2} +
\arctan \!\left( \frac{{\mbox{Re}}\, y(\frac{\pi}{2})}
{{\mbox{Im}}\, y(\frac{\pi}{2})} \right) + n \pi \label{ps1} \\
\varphi^-_n \Big|_0^{\frac{\pi}{2}} &=& (n+1) \pi \label{ps2}
\end{eqnarray}
(we use the usual convention that the arc tangent takes values in~$(-\frac{\pi}{2}, \frac{\pi}{2})$).
Using~(\ref{5B}) these boundary conditions can be expressed purely in terms of~$y$
and the integral of the imaginary part of~$y$.

For the gap estimates we need to control how~$y$ depends on~$\lambda$.
To this end, we differentiate through the complex Riccati equation~(\ref{5c})
and use that $\partial_\lambda V = -1$ according to~(\ref{Vdef})
and~(\ref{mudef}). This gives the linear ODE
\[ y_\lambda' \;=\; -1 - 2 y y_\lambda \;, \]
where the $\lambda$-derivative is denoted by a subscript. This
equation can immediately be integrated using variation of constants.
Applying~(\ref{yint}), we obtain
\begin{equation} \label{ylam}
\left. z^2\: y_\lambda \right|_u^v \;=\; - \int_u^v z^2 \:.
\end{equation}
Substituting the integration-by-parts formula
\[ \int_u^v z^2 \;=\; \int_u^v \frac{1}{2y}\:(z^2)'
\;=\; \left. \frac{z^2}{2y} \right|_u^v +
\int_u^v \frac{V-y^2}{2y^2}\: z^2 \:, \]
we obtain the identity
\begin{equation} \label{ylam2}
z^2\: y_\lambda \Big|_u^v
\;=\; \left. - \frac{z^2}{2y} \right|_u^v
-\int_u^v \frac{V-y^2}{2y^2}\;z^2\:.
\end{equation}
In our estimates we will work both with~(\ref{ylam}) and~(\ref{ylam2}).

\section{Invariant Disk Estimates}
\setcounter{equation}{0}
\label{sec4}
In this section we describe estimates for the complex Riccati
equation~(\ref{5c}) with initial conditions at $u=0$,
\begin{equation} \label{cbvp}
y' \;=\; V - y^2\:,\spc y(0) \;=\; y_0
\end{equation}
on the interval~$[0, \umax)$ with $\umax \in \R^+ \cup \{\infty\}$.
In what follows, the potential~$V \in C^\infty([0, \umax))$
can be any real and smooth function.
The next lemma is the key to all the estimates in this section.
\begin{Lemma} \label{lemmainv}
Let~$\alpha$ be a real function on $[0,\umax)$ which is continuous and
piecewise $C^1$. For a constant $T_0 \geq 1$ we introduce
the functions~$\sigma$, $U$ and~$T$ by
\begin{eqnarray}
\sigma(u) &=& \exp \left( 2 \int_0^u \alpha \right) \label{1g} \\
U(u) &=& V  - \alpha^2 - \alpha' \label{1h} \\
T(u) &=& T_0\: \exp \left( \frac{1}{2}\: {\mbox{TV}}_{[0,u)}
\log |\sigma^2 U| \right) . \label{1i}
\end{eqnarray}
Furthermore, we define the functions $\beta$, $R$ and $m$ by
\begin{eqnarray}
\beta &=& \frac{\sqrt{|U|}}{2} \left(T + \frac{1}{T} \right) \label{1a} \\
R &=& \frac{\sqrt{|U|}}{2} \left(T - \frac{1}{T} \right) \label{1b} \\
m &=& \alpha + i \beta\:. \label{1c}
\end{eqnarray}
Suppose that $U \leq 0$ on~$[0, \umax)$. If a solution $y$ of the boundary value
problem~(\ref{cbvp}) satisfies at $u=0$ the condition
\begin{equation} \label{invest}
|y-m| \;\leq\; R \:,
\end{equation}
then this condition holds for all $u \in [0, \umax)$.
\end{Lemma}
Before coming to the proof, we briefly discuss the statement of this lemma.
If~$\alpha$ is a real solution of the Riccati equation, the function~$U$
as given by~(\ref{1h}) vanishes identically, and thus
$\beta \equiv 0 \equiv R$. In this case, the above lemma reduces to the
trivial statement that $y(0)=\alpha$ implies that $y=\alpha$ on~$[0, \umax)$.
It is more interesting to consider the case that~$\alpha = {\mbox{Re}}\, y$
with~$y$ a {\em{complex}} solution of the Riccati equation. Then
\[ U \;=\; {\mbox{Re}} \left(V - \alpha^2 - \alpha' \right)
\;=\; {\mbox{Re}} \left(V - y^2 - y' \right) - ({\mbox{Im}}\, y)^2
\;=\; -({\mbox{Im}}\, y)^2 \;<\; 0 \:. \]
Moreover, from~(\ref{5A}) we can immediately compute~$\sigma$,
\[ \sigma(u) \;=\; \exp \left(2 \int_0^u {\mbox{Re}}\, y \right)
\;=\; \frac{\rho^2(u)}{\rho^2(0)}\:. \]
Substituting these relations into~(\ref{5C}) yields
\[ \sigma^2 U \;=\; -\frac{\rho^4(u)}{\rho^4(0)} \: ({\mbox{Im}}\, y)^2
\;=\; -\frac{w^2}{\rho^4(0)} \:.\]
Hence the function~$\log |\rho^2 U|$ is a constant, and its total variation
in~(\ref{1i}) vanishes. This means that~$T$ is a constant, and thus
$\beta$ and $R$ are constant multiples of~${\mbox{Im}}\, y$.
Our Lemma states that the circles of radius~$R(u)$ around the
point~$m(u)=\alpha(u)+i \beta(u)$ are invariant under the flow of the Riccati
equation.

If no solution of the Riccati equation is known (and this will of course be the
usual situation), one can take for~$\alpha$ the real part of an
{\em{approximate solution}} of the complex Riccati equation.
In this case, the function~$\log |\rho^2 U|$ will not be constant, but
we can hope that its total variation is small. If this is the case,
our lemma gives an ``improved approximative solution''~$m$ together
with a rigorous error estimate~$R$.
A good candidate for an approximate solution would be the usual wave function
obtained by ``gluing together'' suitable WKB wave functions and Airy
functions as used in the semi-classical analysis of
one-dimensional Schr{\"o}dinger problems.
We remark that the above lemma might even be useful for getting rigorous
error estimates for {\em{numerical}} solutions of the
Sturm-Liouville or Riccati equations.
In this case, one would have to estimate the total variation of~$\log |\rho^2 U|$
from above, and this might be doable numerically if one has some
control of the accuracy of the numerical calculation.
\\[.5em]
{\em{Proof of Lemma~\ref{lemmainv}.}} For $\varepsilon>0$ we set
\beq \label{Teps}
T_\varepsilon(u) \;=\; T_0\: \exp \left( \frac{1}{2}
\int_0^u \left| \frac{|\sigma^2 U|'}{|\sigma^2 U|} \right| + \varepsilon e^{-u}
\right)
\eeq
and let $R_\varepsilon$ and $m_\varepsilon$ be the
functions obtained from~(\ref{1a})--(\ref{1c}) if one replaces
$T$ by $T_\varepsilon$. Since $T_\varepsilon(0)=T(0)$ and
$\lim_{\varepsilon \searrow 0} T_\varepsilon = T$, it suffices to show
that for all $\varepsilon>0$ the following statement holds,
\[ |y-m_\varepsilon|(0) \leq R_\varepsilon(0) \quad \Longrightarrow \quad
|y-m_\varepsilon|(u) \leq R_\varepsilon(u)\;
{\mbox{ for all $u \in [0, \umax)$}}. \]
In order to prove this statement, we will show that the assumption
\beq \label{1d}
|y-m_\varepsilon|(u) \;=\; R_\varepsilon(u)
\eeq
implies that
\beq \label{1k}
|y-m_\varepsilon|'(u) \;<\; R_\varepsilon'(u)\:.
\eeq
In what follows we will often omit the subscript $\varepsilon$.

Assume that~(\ref{1d}) holds and that $U \leq 0$. Then we can represent
$y$ as
\beq \label{1e}
y \;=\; m + R e^{i \varphi}
\eeq
with $\varphi \in [0, 2\pi)$. Furthermore, it follows immediately
from~(\ref{1a}), (\ref{1b}), and~(\ref{1h}) that
\beq \label{1f}
R^2 \;=\; U + \beta^2
\;=\; V -\alpha^2 + \beta^2 - \alpha'\:.
\eeq
Using the above relations together with~(\ref{5c}), we obtain
\begin{eqnarray*}
\lefteqn{ \frac{1}{2}\: \frac{d}{du}|y-m|^2 \;=\;
({\mbox{Re}}\, y - \alpha)\: ({\mbox{Re}}\, y - \alpha)'
+ ({\mbox{Im}}\, y - \beta)\: ({\mbox{Im}}\, y - \beta)' } \\
&\stackrel{(\ref{5c})}{=}&
({\mbox{Re}}\, y - \alpha) \left[V  - ({\mbox{Re}}\, y)^2
+ ({\mbox{Im}}\, y)^2 - \alpha' \right] - ({\mbox{Im}}\, y - \beta)
\left[ 2\: {\mbox{Re}}\, y \:
{\mbox{Im}}\, y + \beta' \right] \\
&=& ({\mbox{Re}}\, y - \alpha) \left[ V  - ({\mbox{Re}}\, y)^2
- ({\mbox{Im}}\, y)^2 +2 \beta\:{\mbox{Im}}\, y - \alpha' \right] \\
&&+ ({\mbox{Re}}\, y - \alpha) \:2 ({\mbox{Im}}\, y - \beta)
\:{\mbox{Im}}\, y \\
&& - ({\mbox{Im}}\, y - \beta)
\left[\beta' + 2 \alpha\: {\mbox{Im}}\, y \right]
\:-\:({\mbox{Im}}\, y - \beta)\:2 ({\mbox{Re}}\, y - \alpha)
\:{\mbox{Im}}\, y \\
&=& ({\mbox{Re}}\, y - \alpha) \left[ V  - ({\mbox{Re}}\, y - \alpha)^2
- ({\mbox{Im}}\, y - \beta)^2 - \alpha^2+\beta^2-\alpha' \right] \\
&& - ({\mbox{Im}}\, y - \beta) \left[\beta' + 2 \alpha \beta \right]
- 2 \alpha \left( ({\mbox{Re}}\, y - \alpha)^2 + ({\mbox{Im}}\, y - \beta)^2
\right) \\
&\stackrel{(\ref{1e})}{=}&
R\: \cos \varphi \left[V - R^2 + |m|^2 - \alpha' - 2 \alpha^2 \right]
-R\: \sin \varphi \left[\beta' + 2 \alpha \beta \right]
-2 \alpha R^2 \\
&\stackrel{(\ref{1f})}{=}&
-2 \alpha R^2 - R\:(\beta' + 2 \alpha \beta)\: \sin \varphi
\;\leq\; -2 \alpha R^2 \:+\: R\:|\beta' + 2 \alpha \beta| \:.
\end{eqnarray*}
Using that $\frac{d}{du} |y-m|^2 = 2 R |y-m|'$, we obtain the simple
inequality
\[ |y-m|' \;\leq\; -2 \alpha R + |\beta' + 2 \alpha \beta| \:. \]
Hence in order to prove~(\ref{1k}), it suffices to show that
\[ R' \;>\; -2 \alpha R \:+\: |\beta' + 2 \alpha \beta| \:. \]

Using~(\ref{1g}), we write the last inequality in the equivalent form
\beq \label{1m}
(\sigma R)' \;>\; |(\sigma \beta)'|\:.
\eeq
In order to prove this inequality, we first use~(\ref{1a}) and~(\ref{1b})
to write the functions $\sigma \beta$ and $\sigma R$ as
\beq \label{1l}
\left. \begin{array}{rcl}
\sigma \beta &=& \displaystyle \frac{1}{2} \left( \sqrt{|\sigma^2 U|}\:T +
\sqrt{|\sigma^2 U|}\:T^{-1} \right) \\[1em]
\sigma R &=& \displaystyle \frac{1}{2} \left( \sqrt{|\sigma^2 U|}\:T -
\sqrt{|\sigma^2 U|}\:T^{-1} \right) . \end{array} \quad \right\}
\eeq
By definition of $T_\varepsilon$~(\ref{Teps}),
\[ \frac{T'}{T} \;=\;
\frac{1}{2} \left| \frac{|\sigma^2 U|'}{|\sigma^2 U|} \right|
+ \varepsilon e^{-u} \:. \]
It follows that
\[ \left\{ \begin{array}{rclcl}
(\sqrt{|\sigma^2 U|}\: T^{-1})' &=&
-\varepsilon e^{-u} \: (\sqrt{|\sigma^2 U|}\: T^{-1})
&\quad& {\mbox{if $|\sigma^2 U|' \geq 0$}} \\[.5em]
(\sqrt{|\sigma^2 U|}\: T)' &=& \varepsilon e^{-u} \: (\sqrt{|\sigma^2 U|}\: T)
&\quad& {\mbox{if $|\sigma^2 U|' < 0$}}\:.
\end{array} \right. \]
Hence when we differentiate through~(\ref{1l}) and set
$\varepsilon=0$, either the first
or the second summand drop out in each equation , and we
obtain $(\sigma R)'=|\sigma \beta|'$. If $\varepsilon>0$, an inspection
of the signs of the additional terms gives~(\ref{1m}).
\QED

The question arises how the function $\alpha$ in the above lemma is to be
chosen. At this point, it is very helpful to regard~(\ref{5ode}) as
the one-dimensional {\em{Schr{\"o}dinger equation}} for a quantum mechanical
wave function~$Y$, because this makes it possible to use ideas
from semi-classical analysis.
In order to explain our method, we first consider the WKB wave
functions~\cite{JWKB}
\[ \phi(u) \;=\; |V|^{-\frac{1}{4}}\: \exp \left( \pm i \int^u \sqrt{|V|} \right) \;, \]
which should be good approximations to fundamental solutions in the
``semiclassical regime''~$V \ll 0$. The corresponding function $y$ is
\[ y(u) \;=\; \frac{\phi'}{\phi} \;=\; \pm i \sqrt{|V|} - \frac{V'}{4 V} \:. \]
Lemma~\ref{lemmainv} should give a good estimate only if $m$ is close
to the exact solution $y$. This leads us to choose for the function
$\alpha = {\mbox{Re}}\, m$ the corresponding expression in the WKB approximation,
\[ \alpha = -\frac{V'}{4 V} \qquad {\mbox{in the ``semiclassical
regime''}}. \]
This gives rise to the following estimate.

\begin{Thm} \label{lemmay1}
Assume that the potential~$V$ is negative and monotone increasing on
the interval~$[0, \umax)$, and that the following condition holds,
\begin{equation} \label{Keq}
K \::=\: \frac{\sup |V''| + {\mbox{\rm{TV}}}\: V''}{\Vmax^2}
+ \sup \frac{V'^2}{|V|^3}
\;\leq\; 1\:,
\end{equation}
where~$\Vmax := \sup V \leq 0$
(and the supremum as well as the total variation are taken on the interval~$[0, \umax)$).
Then the solution~$y$ of the boundary value problem~(\ref{cbvp}) with
initial condition
\begin{equation} \label{yic}
y_0 \;=\; i \sqrt{|V(0)|} - \frac{V'(0)}{4V(0)}
\end{equation}
satisfies on~$[0,\umax)$ the inequalities
\begin{eqnarray}
\left| y - i \sqrt{|V|} + \frac{V'}{4V} \right| &\leq& 20\: \sqrt{|V|} \:K  \label{A} \\
{\mbox{\rm{Im}}}\, y &\geq& \frac{\sqrt{|V|}}{10} \:. \label{B}
\end{eqnarray}
\end{Thm}
{\Proof}
We introduce on $[0,\umax]$ the function $\alpha$ by
\[ \alpha(u) \;=\; -\frac{V'}{4V} \:. \]
Then from~(\ref{1h}),
\begin{eqnarray}
\alpha' &=& -\frac{V''}{4V} + \frac{V'^2}{4 \:V^2} \\
U &=& V \left(1 + \frac{V''}{4 V^2} \:+\: \frac{5\:V'^2}{16\:|V|^3}
\right) \:. \label{UVeq}
\end{eqnarray}
Using the inequality~(\ref{Keq}) we get
\begin{equation} \label{UV}
2V \;\leq\; U \;\leq\; \frac{V}{2}\:.
\end{equation}
In particular, $U$ is negative.

The inequalities~(\ref{UVeq}) and~(\ref{Keq}) allow us to estimate
$\sqrt{|U|} - \sqrt{|V|}$,
\begin{equation}
\left| \sqrt{|U|} - \sqrt{|V|} \right| \;=\; \frac{|U-V|}{\sqrt{|U|} + \sqrt{|V|}}
\;\leq\; \sqrt{|V|}\:\left| \frac{U-V}{V} \right|
\;\leq\; \frac{\sqrt{|V|}}{2}\: K \:. \label{UmV}
\end{equation}
Dividing by~$\sqrt{|V|}$ and $\sqrt{|U|}$, we obtain furthermore
\begin{equation} \label{UdV}
\frac{1}{1+K} \;\leq\; \sqrt{\frac{|V|}{|U|}} \;\leq\; 1+K \:.
\end{equation}

Choosing~$T_0=1+K$, we have the following estimates at $u=0$,
\begin{eqnarray*}
|y-m| &=& |\sqrt{|V|} - \beta|
\;=\; \left| \sqrt{|V|} - \frac{\sqrt{|U|}}{2}
\left((1+K) + \frac{1}{1+K} \right) \right| \\
&=& \frac{\sqrt{|U|}}{2} \;\left| (1+K) + \frac{1}{1+K} - 2 \:\sqrt{\frac{|V|}{|U|}} \right| .
\end{eqnarray*}
Applying~(\ref{UdV}) we obtain
\begin{eqnarray*}
|y-m| &\leq& \frac{\sqrt{|U|}}{2} \left( (1+K) - \frac{1}{1+K} \right)
\;=\; R \:.
\end{eqnarray*}
We conclude that the inequality~(\ref{invest}) holds at~$u=0$.

Hence we can apply Lemma~\ref{lemmainv} and obtain that~(\ref{invest})
holds for all $u \in [0,\umax)$. Combining this with the
inequalities~(\ref{UV}) and~(\ref{UdV}) we obtain
\begin{eqnarray}
\lefteqn{ \hspace*{-1cm} \left|y-i \sqrt{|V|} + \frac{V'}{4V} \right| \;\leq\;
|y-m| + |\beta-\sqrt{|V|}| \;\leq\; R + |\beta-\sqrt{|V|}| } \nonumber \\
&=& R + \frac{\sqrt{|U|}}{2} \left( T + \frac{1}{T} - 2 \sqrt{\frac{|V|}{|U|}} \right)
\;=\; \sqrt{|U|} \left(T - \sqrt{\frac{|V|}{|U|}} \right) \nonumber \\
&\leq& 2 \sqrt{|V|} \left(T - \frac{1}{1+K} \right)
\;\leq\; 2 \sqrt{|V|} \:\left(T - 1 + K \right) \label{e1} \\
{\mbox{Im}}\, y &\geq& \beta - R \;=\; \frac{\sqrt{|U|}}{T}
\;\geq\; \frac{\sqrt{|V|}}{\sqrt{2} T} \:.
\label{e2}
\end{eqnarray}
It remains to estimate the function $T$, (\ref{1i}). We first
compute~$\sigma$ and $\sigma^2 U$,
\begin{eqnarray}
\sigma &=& \sqrt{ \frac{V_0}{V(u)} } \\
\frac{\sigma^2 U}{|V_0|} &=&
-1 - \frac{V''}{4 V^2} \:-\: \frac{5\:V'^2}{16\:|V|^3} \:, \label{rU}
\end{eqnarray}
where we set $V_0=V(0)$.
Applying~(\ref{Keq}) we immediately obtain the inequalities
\[ \frac{1}{2} \;\leq\; \left| \frac{\sigma^2\:U}{V_0} \right| \;\leq\; 2\:. \]
The lower bound allows us to leave out the logarithm in the total
variation in the definition of $T$; namely,
\[ {\mbox{TV}}_{[0,u)} \log |\sigma^2 U| \;=\;
\int_0^u \left| \frac{(\sigma^2 U)'}{\sigma^2 U} \right|
\;\leq\; 2 \int_0^u \left| \frac{(\sigma^2 U)'}{|V_0|} \right|
\;=\; 2\:{\mbox{TV}}_{[0,u)} \frac{\sigma^2 U}{|V_0|}\:. \]
We substitute~(\ref{rU}) into this equation and estimate the total
variation of the individual terms using~(\ref{Keq}) as well as
the monotonicity of $V$,
\begin{eqnarray*}
{\mbox{TV}}_{[0, u)}\: \frac{V''}{V^2} &\leq&
\int_0^u \frac{|V'''|}{V^2} + 2 \int_0^u
\frac{|V''|\: V'}{(-V)^3}
\;\leq\; \frac{{\mbox{TV}}\, V'' + \sup |V''|}{\Vmax^2} \\
{\mbox{TV}}_{[0, u)}\: \frac{V'^2}{(|V|)^3} &\leq&
\int_0^u \frac{2\:|V''| \: V'}{(|V|)^3} +
\int_0^u \frac{3\:|V'|^3}{V^4}
\;\leq\; \frac{\sup |V''|}{\Vmax^2}
+ \int_0^u \frac{3\:|V'|^3}{V^4} \:.
\end{eqnarray*}
In the last term we can integrate by parts,
\[ \int_0^u \frac{3\:|V'|^3}{V^4} \;=\;
\int_0^u V'^2 \left( (-V)^{-3} \right)'
\;=\; \left. \frac{V'^2}{|V|^3} \right|_0^u
- \int_0^u \frac{2 V''\: V'}{|V|^3}
\;\leq\; \sup \frac{V'^2}{|V|^3} + \frac{\sup |V''|}{\Vmax^2} \:. \]
Collecting all the terms and using~(\ref{Keq}) we conclude that
\[ {\mbox{TV}}_{[0, u)} \log |\sigma^2 U| \;\leq\; 2 K\:. \]
We substitute this bound into~(\ref{1i}) and use that $T_0=1+K$
to obtain the bound
\[ T-1 \;=\; (1+K)\:e^K-1 \;\leq\;
|e^K-1| + K e^K \;\leq\; 2e\:K \:.  \]
Using this bound in~(\ref{e1}) and~(\ref{e2}) concludes the proof.
\QED

The condition~(\ref{Keq}) will clearly be violated when $|V|$ becomes
small. This is not astonishing because the WKB approximation fails
near the zeros of the potential. In this ``quantum regime'', there is no
canonical candidate for~$\alpha$, and therefore we simply take
\[ \alpha = {\mbox{const}} \qquad {\mbox{in the ``quantum regime''}}. \]
We state the corresponding estimate in such a way that it can
easily be ``pasted together'' with the result of Lemma~\ref{lemmay1}.
\begin{Thm} \label{lemmay2}
Assume that the potential~$V$ is negative and monotone (increasing
or decreasing) on $[0, \umax)$, and that for some
constant~$\kappa>0$ the following condition holds,
\begin{equation} \label{V0cond}
\sqrt{|V_0|}\; \umax \;\leq\; \kappa
\end{equation}
(with $V_0=V(0)$).
Then any solution~$y$ of the boundary value problem~(\ref{cbvp})
which is bounded by
\[ |y_0| \;\leq\; c_1\:\sqrt{|V_0|} \;,\spc
{\mbox{\rm{Im}}}\,y_0 \;\geq\; \frac{\sqrt{|V_0|}}{c_1} \]
for a suitable constant~$c_1 \geq 1$ satisfies on $[0,\umax)$ the inequalities
\[ |y| \;\leq\; c_2\:\sqrt{\|V\|_\infty} \;,\spc
{\mbox{\rm{Im}}}\,y \;\geq\; \frac{1}{c_2}\: \frac{|V_0|}{\sqrt{\|V\|_\infty}} \]
where~$\|V\|_\infty:= \sup_{[0, \umax)} |V|$ and~$c_2$
is a constant which depends only on $\kappa$ and $c_1$.
\end{Thm}
{\Proof} Let~$\alpha$ be the constant function~$\alpha=\sqrt{|V_0|}$.
Then the function $U=V-\alpha^2$ is clearly negative. A simple calculation
shows that by choosing~$T_0 = 2 \:c_1 (1+c_1)^2$, we can arrange that
$|y_0-m(0)|\leq R(0)$. Lemma~\ref{lemmainv} yields that
$|y-m| \leq R$ for all $u \in [0, \umax)$.

Since~$\alpha$ is a constant, the function~$\sigma$ is given by~$\sigma(u) = e^{2 \alpha u}$
and thus
\[ |\sigma^2 U| \;=\; e^{4 \alpha u} \left( \alpha^2-V \right) \:. \]
As a consequence,
\[ \frac{|\sigma^2 U|'}{|\sigma^2 U|} \;\leq\; 4 \alpha - \frac{V'}{\alpha^2-V}\:. \]
If we integrate and use~(\ref{V0cond}), we obtain the following bound for $T$,
\begin{eqnarray*}
T &\leq& T_0 \: e^{2 \alpha u}
\left( \frac{\sqrt{\alpha^2-V_0}}{\sqrt{\alpha^2 - V}} + \frac{\sqrt{\alpha^2-V}}{\sqrt{\alpha^2 - V_0}}
\right) \\
&\leq& T_0  \: e^{2 \alpha u}\; \frac{4\sqrt{\|V\|_\infty}}{\alpha}
\;\leq\; 4 T_0\: e^{2 \kappa} \;\sqrt{\frac{\|V\|_\infty}{|V_0|}}  =: T_{max} \:.
\end{eqnarray*}

Finally, we bound $y$ by
\begin{eqnarray*}
|y| &\leq& |y-m| + |m| \;\leq\; R + \alpha + \beta \\
&=& \sqrt{|U|}\:T + \alpha \;\leq\; (2T+1)\: \alpha \\
{\mbox{Im}}\, y &\geq& \beta-R \;=\; \frac{\sqrt{|U|}}{T}
\;\geq\; \frac{\alpha}{T}\:.
\end{eqnarray*}
These are the desired inequalities if we set
$c_2=2T_{max}+1 = 8 c_1 \:(1+c_1)^2\: e^{2 \kappa}+1$.
\QED

It is obvious from~(\ref{Vdef}, \ref{mudef}) that the potential~$V$ has a singularity
at~$u=0$. We now explain how Lemma~\ref{lemmainv}
can be used for estimates near such a singular point.
We will restrict attention to the case~$k=0$, but our method applies
similarly to general~$k$. In order to find a good candidate for the
function~$\alpha$, we consider on the interval $[0, \umax)$
the Sturm-Liouville equation with a potential which at $u=\umax$
has the same singular behavior as~(\ref{Vdef}),
\begin{equation} \label{schexact}
z''(u) \;=\; -\frac{1}{4 \:(\umax-u)^2}\: z \:.
\end{equation}
Setting $v=\umax-u$, this differential equation has the two
fundamental solutions $\phi_1=\sqrt{v}$ and $\phi_2=\sqrt{v}\:\log v$,\
and therefore the function
\[ z \;=\; \sqrt{v} \left( 1 - i \log v \right) \]
is a complex solution. The corresponding solution of the complex
Riccati equation is given by
\begin{equation} \label{yexact}
y \;=\; \frac{z'}{z} \;=\; -\frac{1}{2 v} + \frac{i}{v\: (1-i \log v)}
\;=\; \left(-\frac{1}{2v} - \frac{\log v}{v\:(1+\log^2 v)} \right)
+ \frac{i}{v\:(1+\log^2 v)} \:.
\end{equation}
Choosing~$\alpha$ equal to the real part of this function
gives rise to the following estimate.

\begin{Lemma} \label{thmpole}
Suppose that the potential~$V$ is on~$[0, \umax)$ of the form
\[ V \;=\; -\frac{1}{4}\:\frac{1}{(\umax-u)^2} + B(u) \]
with a function~$B$ which is monotone (decreasing or increasing)
and satisfies the inequality
\begin{equation} \label{Bcond}
\umax^2\:(1+\log^2 \umax)^2\: \|B\|_\infty \;\leq\; \frac{1}{8}
\end{equation}
(with~$\|B\|_\infty := \sup_{[0, \umax)} |B|$).
Then any solution~$y$ of the boundary value problem~(\ref{cbvp})
with initial conditions bounded by
\[ |y_0| \;\leq\; C\:\sqrt{|V_0|} \;,\spc
{\mbox{\rm{Im}}}\,y_0 \;\geq\; \frac{\sqrt{|V_0|}}{C} \]
for any constant~$C \geq 1$ satisfies on $[0,\umax)$ the inequalities
\begin{eqnarray}
|y| &\leq& \frac{64\:C^3}{\umax-u} \\
{\mbox{\rm{Im}}}\,y &\leq& 64\:C^3\: (1+\log^2 \umax)\;
\frac{1}{(\umax-u)\: \log^2(\umax-u)} \label{Imy} \\
{\mbox{\rm{Im}}}\,y &\geq& \frac{1}{64\:C^3\: (1+\log^2 \umax)}\;
\frac{1}{(\umax-u)\: \log^2(\umax-u)} \:.
\end{eqnarray}
\end{Lemma}
{\Proof} We set $v=\umax-u$ and choose for $\alpha$ the real function
\[ \alpha \;=\; -\frac{1}{2v} - \frac{\log v}{v\:(1+\log^2 v)}\:. \]
Using that~$\alpha = {\mbox{Re}}\, y$ with~$y$ according to~(\ref{yexact})
and that~$y$ is a solution of the complex Riccati equation corresponding
to the Sturm-Liouville equation~(\ref{schexact}), we obtain
\begin{equation}
U \;=\; V - \alpha^2 - \alpha' \;=\;
{\mbox{Re}} \left( V - y^2 - y' \right) - ({\mbox{Im}} \,y)^2 \\
\;=\; B - \frac{1}{v^2\:(1+\log^2 v)^2} \:.
\end{equation}
Using the assumption~(\ref{Bcond}) together with the fact that the
function~$v^2(1+\log^2v)$ is monotone increasing, we obtain
that~$U$ is negative.

At~$u=0$, the potentials~$V$ and $U$ can easily be bounded from above
and below,
\begin{eqnarray*}
-\frac{3}{2} &\leq& -1-4v^2 |B| \;\leq\;l 4v^2\:V \;=\; -1 + 4 v^2 |B| \;\leq\; -\frac{1}{2} \\
-\frac{3}{2} &\leq& 4v^2\:(1+\log^2 v)^2\: U \;=\; -1 + v^2 \:(1+\log^2 v)^2\: B \;\leq\; -\frac{1}{2}
\end{eqnarray*}
and in particular
\[ \frac{1}{2} \;\leq\; \frac{(1+\log^2 v)\: \sqrt{|U|}}{|\sqrt{V}|} \;\leq\; 2 \:. \]
A simple calculation shows that by choosing
$T_0 = 2 C (1+C)^2 \:(1+\log^2 \umax)$, we can arrange that
$|y_0-m(0)|\leq R(0)$. Lemma~\ref{lemmainv} yields that
$|y-m| \leq R$ for all $u \in [0, \umax)$.

Writing the function~$\alpha$ in the form
\[ \alpha \;=\; \frac{d}{du} \log \left( \sqrt{v\:(1+ \log^2 v)} \right) \, \]
we can immediately compute $\sigma^2 U$,
\begin{eqnarray*}
\sigma^2 &=& v^2\:(1+\log^2 v)^2 \\
|\sigma^2 \:U| &=& 1 - v^2\:(1+\log^2 v)^2\:B \:.
\end{eqnarray*}
Using the bound~(\ref{Bcond}), we obtain
\begin{equation}
{\mbox{TV}}_{[0,u)} \log |\sigma^2\:U| \;\leq\;
2 \:{\mbox{TV}}_{[0,u)} |\sigma^2\:U| \;\leq\;
4\: \umax^2\:(1+\log^2 \umax)^2\: \|B\|_\infty \;\leq\; 2 \:,
\end{equation}
and thus $T$ is bounded by~$T \leq T_0 e^2 \leq
64 C^3 \:(1+\log^2 \umax)$. Finally, we combine
the above estimates with the inequalities
\[ |y| \;\leq\; R + |\alpha|+\beta \;,\spc
R-\beta \;\leq\; {\mbox{Im}}\, y \;\leq\; R+\beta\:. \]

\vspace*{-.75cm}
\QED
The estimate~(\ref{Imy}) is very useful because it
shows that the pole of~${\mbox{Im}}\, y$ at~$u=0$ is integrable.

\section{Convexity Estimates}
\setcounter{equation}{0} \label{sec5}
The estimates of the previous section gave us good control
of the solutions of the boundary value problem~(\ref{cbvp})
provided that the potential is negative.
In this section we proceed with estimates in the case
that~$V$ is {\em{positive}}, $V \geq 0$.
Under this assumption, it is a simple observation
that~$\rho^2$ is convex, because
\begin{equation} \label{convex}
(\rho^2)'' \;=\; (\overline{z} z)'' \;=\; 2\:(V + |y|^2)\: \rho^2 \;>\; 0 \:.
\end{equation}
This fact will be essential for the estimates in this section.

We begin with a lemma which bounds~$\rho$ from below.
\begin{Lemma} \label{lemmaconvex1}
Suppose that~$V$ is positive and monotone increasing
on~$[0, \umax)$. Then every solution of the boundary value problem~(\ref{cbvp})
satisfies on~$[0, \umax)$ the inequality
\[ \rho \;\geq\; \rho_0\: \frac{{\mbox{\rm{Im}}}\, y_0}{|y_0|} \]
(with $\rho_0 = |z(0)|$ and $z$ any solution of the corresponding
complex Sturm-Liouville equation~(\ref{5csch})).
\end{Lemma}
{\Proof}
Differentiating the equation $\rho'=\rho\: {\mbox{\rm{Re}}}(y)$ and
using the complex Riccati equation~(\ref{5c}), we get
\[ \rho'' \;=\; \rho\: ({\mbox{\rm{Re}}}\,y)^2 + \rho\: {\mbox{\rm{Re}}}(V-y^2)
\;=\; \left(V + ({\mbox{Im}}\, y)^2 \right) \rho \:, \]
and using~(\ref{5C}) we obtain the following differential equation
for $\rho$,
\begin{equation} \label{rhoeq}
\rho'' \;=\; V \rho + \frac{w^2}{\rho^3}\:.
\end{equation}

We let~$\underline{\rho}(u)$ be the solution of the boundary value problem
\begin{equation} \label{bvp}
\underline{\rho}'' = \frac{w^2}{\underline{\rho}^3} \spc {\mbox{with}} \spc
\underline{\rho}(0)=\rho_0\:,\quad \underline{\rho}'(0)=\rho'(0) \equiv \rho_0\:
{\mbox{\rm{Re}}}\, y_0 \:.
\end{equation}
The function~$\underline{\rho}$ is a solution of~(\ref{rhoeq}) in the case~$V \equiv 0$.
Therefore, $\underline{\rho}$ can be written explicitly in the form $\underline{\rho}=|
\underline{z}|$ with
$\underline{z}$ a solution of the complex Sturm-Liouville equation without potential with Wronskian equal to~$w$, i.e.
\[ \underline{z}'' = 0 \spc{\mbox{and}}\spc {\mbox{\rm{Im}}} (\overline{\underline{z}}\:
\underline{z}') \;=\;
w \;=\; \rho_0^2\: {\mbox{\rm{Im}}}\, y_0\:. \]
A short calculation shows that~$\underline{\rho}$
has the simple form
\[ \underline{\rho}(u) \;=\; \rho_0\:\left| 1 + y_0\: u \right| . \]
This function is defined even for all~$u \in \R$. It is convex,
and its minimum is computed to be
\beq \label{rholb}
\min_{u \in \R} \underline{\rho}(u) \;=\; \rho_0\: \frac{{\mbox{\rm{Im}}}\, y_0}{|y_0|}
\:.
\eeq

We introduce the set~$I \subset \R^2$ by
\[ I \;=\; \left\{ \left(\underline{\rho}(x),
[\underline{\rho}'(x), \infty) \right) {\mbox{ with }} x \in \R \right\} \:. \]
Let us show that~$I$ is an invariant region for~$\rho$ in phase space, i.e.\ that for all~$u \in [0, \umax)$,
\beq \label{invcond}
(\rho(u), \rho'(u)) \in I \:.
\eeq
Once this is shown, the lemma follows immediately from~(\ref{rholb}).
Due to our initial condition, (\ref{invcond}) is
clearly satisfied at~$u=0$. Thus assume that~$[0, v]$ with~$0 \leq v<\umax$
is the maximal interval where~(\ref{invcond}) holds.
Then the point~$(\rho(v), \rho'(v))$ lies on the boundary of~$I$.
Using that
$\underline{\rho}$ is convex and thus~$\underline{\rho}'$ is
monotone increasing, one finds that either
\[ \rho(v) \;=\; \min \underline{\rho} \quad {\mbox{and}} \quad
\rho'(v) \;>\; 0 \]
or else there is~$x \in \R$ such that
\[ \rho(v) \;=\; \underline{\rho}(x) \quad {\mbox{and}} \quad
\rho'(v) \;=\; \underline{\rho}'(x) \:. \]
In the first case, it is obvious that the gradient of~$(\rho(v),
\rho'(v))$ is pointed towards the interior of~$I$.
In the second case, the estimate
\[ \rho''(v) \;=\; V\, \rho(v) + \frac{w}{\rho(v)^3} \;>\;
\frac{w}{\rho(v)^3} \;=\; \frac{w}{\underline{\rho}(x)^3} \;=\;
\underline{\rho}''(x) \]
yields that
\[ \rho'(v) \;=\; \underline{\rho}'(x) \quad {\mbox{and}} \quad
\rho''(v) \;>\; \underline{\rho}''(x)\:. \]
Hence the gradient of~$(\rho(u), \rho'(u))$ again points towards the
interior of~$I$. We conclude that~(\ref{invcond}) holds also in
an interval~$[v,v+\varepsilon)$ with~$\varepsilon>0$, a contradiction.
\QED
This lemma has the following immediate consequence.
Due to the convexity of $\rho$,
\[ \sup_{[0,u)} \rho \;\leq\; \rho_0 + \rho(u)
\;=\; \rho(u) \left( 1 + \frac{\rho_0}{\rho(u)} \right)
\;\leq\; \rho(u) \left( 1 + \frac{|y_0|}{{\mbox{\rm{Im}}}\, y_0} \right) \]
and hence
\begin{equation}
\sup_{[0,u)} \rho \;\leq\; \rho(u) \;\frac{2 |y_0|}{{\mbox{\rm{Im}}}\, y_0}\:.
\end{equation}

In regions where the potential~$V$ is large, we expect that~$\rho$ should
increase exponentially.
The next lemma quantifies this exponential increase of $\rho$ by
showing that in the ``semiclassical regime'' $V \gg 0$, the integral over~$\rho^2$ is
much smaller than the supremum of~$\rho^2$.

\begin{Lemma} \label{lemmaconvex2}
Suppose that~$V$ is positive and monotone increasing
on~$[0, \umax)$. Then every solution of the boundary value problem~(\ref{cbvp})
satisfies on~$[0, \umax)$ the inequality
\[ \int_0^u \rho^2 \;\leq\; L \:\sup_{[0,u)} \rho^2 \]
with~$L$ given by
\begin{equation} \label{Leq}
L \;=\;
\sup \left(\frac{3}{\sqrt{V}} + \frac{V'}{V^2} \right)
+ {\mbox{\rm{TV}}} \frac{V'}{V^2} .
\end{equation}
\end{Lemma}
{\Proof} We substitute the differential equation for~$\rho^2$, (\ref{convex}),
into the integral,
\[ \int_0^u \rho^2 \;=\; \frac{1}{2} \int_0^u \frac{1}{V+|y|^2}\:
(\rho^2)'' \:. \]
Integrating by parts gives
\[ \int_0^u \rho^2 \;=\; \left. \frac{(\rho^2)'}{2 \:(V+|y|^2)} \right|_0^u
\:-\: \frac{1}{2} \int_0^u \left( \frac{1}{V+|y|^2} \right)' \:
(\rho^2)' \:. \]
Using the estimates
\begin{eqnarray*}
\left| \frac{(\rho^2)'}{V+|y|^2} \right| &\leq&
\frac{2 \rho^2 \:|y|}{V+|y|^2} \;\leq\; \frac{2 \rho^2 \:|y|}{2 \sqrt{V}\:|y|}
\;=\; \frac{\rho^2}{\sqrt{V}} \\
\left| \left( \frac{1}{V+|y|^2} \right)' \right| &\leq&
\frac{V' + 2 |y|\: |V-y^2|}{(V+|y|^2)^2} \;\leq\;
\frac{V'}{V^2} + \frac{2 |y|}{V + |y|^2} \;\leq\;
\frac{V'}{V^2} + \frac{1}{\sqrt{V}}
\end{eqnarray*}
we obtain
\[ \int_0^u \rho^2 \;\leq\; \sup_{[0,u)} \frac{\rho^2}{\sqrt{V}}
\:+\: \frac{1}{2} \int_0^u \left(\frac{V'}{V^2} + \frac{1}{\sqrt{V}} \right)
\left|(\rho^2)' \right|\:. \]
When integrating by parts once again we must be careful because the function
$(\rho^2)'$ may change signs. However, since $\rho^2$ is convex, it changes
signs at most once, and therefore we get positive boundary terms at most twice,
\[ \int_0^u \left(\frac{V'}{V^2} + \frac{1}{\sqrt{V}} \right)
\left|\frac{d}{du} \rho^2 \right| \;\leq\;
2 \sup_{[0,u)} \left(\rho^2\:\frac{V'}{V^2} + \frac{\rho^2}{\sqrt{V}} \right)
\:+\: \int_0^u \left| \left(\frac{V'}{V^2} + \frac{1}{\sqrt{V}} \right)' \right|
\rho^2 \:. \]
Finally, we can estimate the last integral by
\begin{eqnarray*}
\int_0^u \left| \left(\frac{V'}{V^2} + \frac{1}{\sqrt{V}} \right)' \right|
\rho^2 &\leq& \sup_{[0,u)} \rho^2\:
{\mbox{\rm{TV}}}_{[0,u)} \left(\frac{V'}{V^2} + \frac{1}{\sqrt{V}} \right) \\
&=& \sup_{[0,u)} \rho^2 \left(
{\mbox{\rm{TV}}}_{[0,u)} \frac{V'}{V^2} \:+\: \sup_{[0,u)} \frac{1}{\sqrt{V}} \right) ,
\end{eqnarray*}
where in the last step we used the monotonicity of~$V$.
\QED

\section{Elementary Properties of the Potential}
\label{sec6}
\begin{figure}[p]
\begin{center}
\input{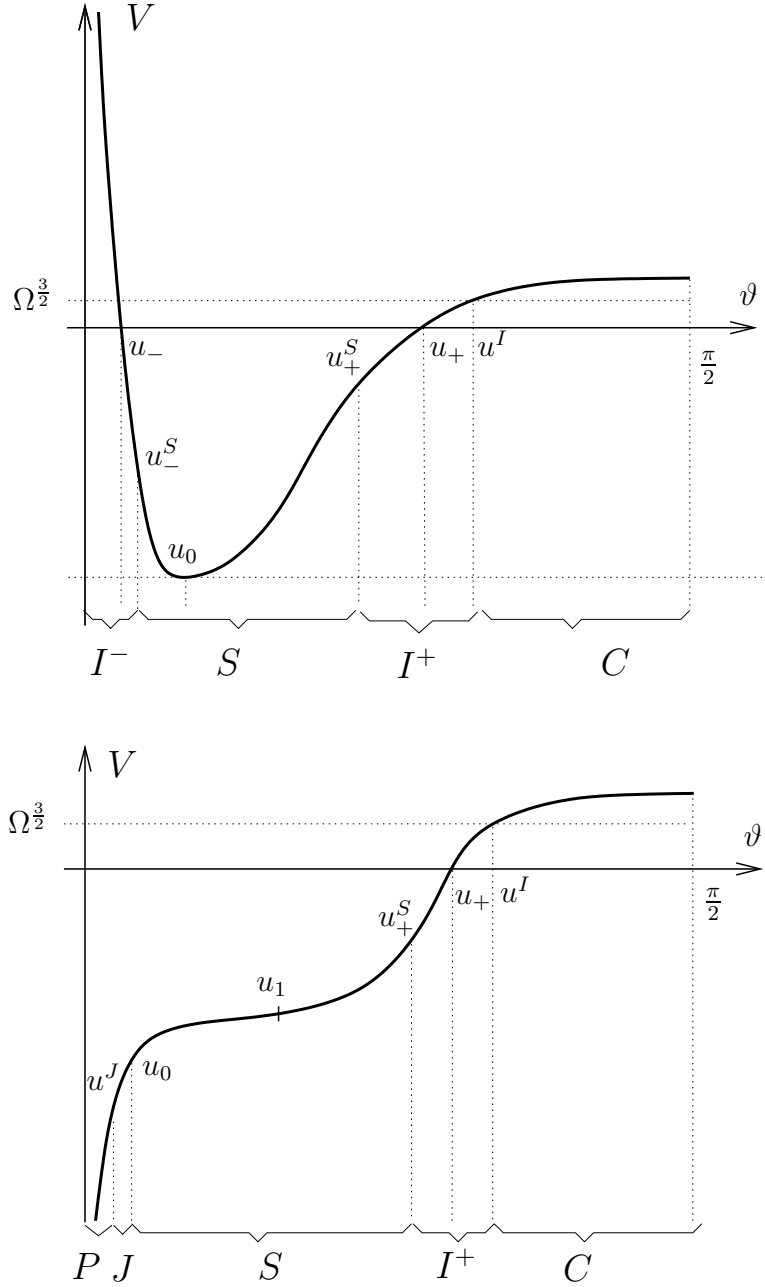}
\caption{The potential~$V$ in the cases $k \neq 0$ (top) and $k=0$ (bottom).}
\label{fig1}
\end{center}
\end{figure}

\setcounter{equation}{0}
In this section we shall analyze the potential~$V$ (\ref{Vdef}, \ref{mudef})
for large~$\lambda$ and~$\Omega$. More precisely, we consider the range
\begin{equation} \label{range}
\Omega > \Omega_0 \spc {\mbox{and}} \spc \lambda \;>\; 2 \Lambda \Omega
\end{equation}
for parameters~$\Omega_0$ and~$\Lambda$, which we can choose as large as we want.
Then the potential looks qualitatively as in Figure~\ref{fig1}.
In the case $k\neq0$, $V$ has a unique minimum $u_0$ given by
\beq \label{u0def}
\sin^2 u_0 \;=\; \frac{1}{\Omega} \:\sqrt{ k^2 - \frac{1}{4} }\:,
\eeq
and the potential is negative at the minimum,
\beq \label{62a}
V(u_0) \;=\; \Omega \left(2\: \sqrt{ k^2 - \frac{1}{4} } + 2k \right)
- \lambda - \frac{1}{4} \;<\; -\Lambda \Omega
\eeq
(where in the last step we possibly increased~$\Lambda$).
$V$ is strictly decreasing on the interval $[0, u_0]$ and tends to
infinity as $u \searrow 0$. Thus there is a
unique $u_- \in [0, u_0]$ with $V(u_-)=0$.
On the interval $[u_0, \frac{\pi}{2}]$, $V$ is strictly increasing.
Thus there is at most one $u_+ \in (u_0, \frac{\pi}{2}]$ with
$V(u_+)=0$. If no such $u_+$ exists, we set $u_+=\frac{\pi}{2}$.
For a given parameter~$\kappa>0$ (which will be specified later)
we set~$\Delta u = \kappa/\sqrt{\Lambda \Omega}$. It is easily verified that
by choosing~$\Lambda$ sufficiently large we can arrange
that~$V(u_0 \pm \Delta u) < 0$. As a consequence,
$|u_\pm-u_0| \geq \Delta u$ and thus, using~(\ref{62a}),
\[ |V(u_0)| \:|u_\pm - u_0|^2 \;>\; \kappa^2 \:. \]
Using monotonicity, we can thus uniquely introduce points~$u_+^S \in (u_0, u_+)$
and~$u_-^S \in (u_-, u_0)$ by the condition that
\begin{equation} \label{usdef}
|V(u_\pm^S)| \:|u_\pm - u_\pm^S|^2 \;=\; \kappa^2 \:.
\end{equation}
Finally, we introduce the point~$u^I \in (u_+, \frac{\pi}{2}]$
by the condition $V(u^I)=\Omega^{\frac{3}{2}}$. If no such point
exists, we set~$u^I=\frac{\pi}{2}$.

In the case~$k=0$, $V$ is monotone increasing on the whole
interval~$(0, \frac{\pi}{2}]$. We set
\beq \label{uJ01def}
u^J \;=\; \frac{1}{8 \sqrt{\lambda}\: \log^2 \lambda} \;,\qquad
u_0 \;=\; \frac{\kappa}{\sqrt{\lambda}} \qquad{\mbox{and}}\qquad
u_1 \;=\; \frac{1}{\sqrt{\Omega}} \:.
\eeq
The points~$u_+$, $u_+^S \in (u_1, u_+)$ and $u^I$ are introduced as in
the case~$k \neq 0$.

We consider on~$(0,\frac{\pi}{2}]$ the solution~$y$ of the
complex Riccati equation~(\ref{5c}) with initial condition
\begin{equation} \label{initc}
y(u_0) \;=\; i \sqrt{|V(u_0)|} - \frac{V'(u_0)}{4 V(u_0)}\:.
\end{equation}
The next lemmas make the following statements precise:
The intervals~$S$ (as introduced in Figure~\ref{fig1})
are ``semiclassical'' in the sense that
Theorem~\ref{lemmay1} applies. On~$C$ we can use the convexity
estimate of Lemma~\ref{lemmaconvex2}. The interval~$P$ near the
pole can be treated by Theorem~\ref{thmpole}. Finally, the
``intermediate regions''~$I_\pm$ and~$J$ can be controlled with
Theorem~\ref{lemmay2} and Lemma~\ref{lemmaconvex1}.

\begin{Lemma} \label{lemmapr1}
For every~$\delta>0$ and~$k \in \Z$ there are parameters
$\kappa, \Lambda, \Omega_0 >0$ such that
for all~$\Omega, \lambda$ in the range~(\ref{range}),
the quantity~$K$ as defined by (\ref{Keq}) is on the interval~$S$ bounded by
\[ K \;\leq\; \delta\:. \]
\end{Lemma}
{\Proof}
The third derivative of $V$ can be written in the form
\[ V''' \;=\; \frac{\cos u}{\sin^5 u} \:
{\mbox{(polynomial in $\sin^2 u$ of degree $3$)}}. \]
Hence $V'''$ has on any interval $[u,v] \subset [0, \frac{\pi}{2}]$ at
most $4$ zeros. Thus, after splitting up $[u,v]$ into at most four
subintervals, $V'''$ has on each subinterval a fixed sign. On any such
subinterval $[\underline{u}, \underline{v}]$ we can apply the estimate
\[ \int_{\underline{u}}^{\underline{v}} |V'''| \:du \;\leq\;
|V''(\underline{u})| + |V''(\underline{v})| \:. \]
This makes it possible to control the total variation of~$V''$
in~(\ref{Keq}) by~$8 \sup |V''|$. We conclude that it
suffices to show that on the interval~$I$ the following two inequalities
hold,
\begin{eqnarray}
\frac{V'^2}{|V|^3} &\leq& \delta \label{Vcd1} \\
\frac{|V''|}{V^2} &\leq& \delta \label{Vcd2} \:.
\end{eqnarray}
We treat three cases separately. \\[.8em]
{\bf{First case:}} $k=0$ and $u_+ \geq \frac{3 \pi}{8}$. \\
On the interval~$[\frac{5 \pi}{16}, \frac{\pi}{2}]$, the potential~$V$
is concave; more precisely,
\beq \label{Vpp}
-\Omega^2 \;\leq\; V'' \;\leq\; -\frac{\Omega^2}{4}\:.
\eeq
Integration yields for all~$\tau \in (\frac{5 \pi}{16}, u_+]$
the following bounds for~$V'$ and~$V$,
\begin{eqnarray}
\frac{\Omega^2}{4} \left(\frac{\pi}{2}-\tau \right) &\leq& V'(\tau) \;\leq\;
\Omega^2 \left(\frac{\pi}{2}-\tau \right) \label{Vp1} \\
V(u_+)-V(\tau) &\geq& \frac{\Omega^2}{8} \left. \left( \frac{\pi}{2} - t \right)^2
\right|_{u_+}^\tau \;=\; \frac{\Omega^2}{8}\: (u_+-\tau)\: (\pi-u_+-\tau) \:. \nonumber
\end{eqnarray}
Since~$V(u_+)$ is either zero or negative, it follows that
\begin{equation} \label{V1}
|V(\tau)| \;\geq\; \frac{\Omega^2}{8}\: (u_+-\tau)\: (\pi-u_+-\tau) \:.
\end{equation}
Combining the inequalities~(\ref{Vpp}, \ref{Vp1}, \ref{V1}), we obtain
for all~$\tau \in (\frac{5 \pi}{16}, u_+^S]$ the estimates
\begin{eqnarray}
\frac{V'(\tau)^2}{|V(\tau)|^3} &\leq&
8^3\: \frac{ \Omega^4 \left(\frac{\pi}{2}-\tau \right)^2 }
{\Omega^6 (u_+-\tau)^3\: (\pi-u_+-\tau)^3} \;\leq\;
\frac{8^3}{\Omega^2\: (u_+ - u_+^S)^4} \label{Vpq} \\
\frac{|V''(\tau)|}{V^2(\tau)} &\leq&
64\: \frac{\Omega^2}{\Omega^4 (u_+-\tau)^2\: (\pi-u_+-\tau)^2}
\;\leq\; \frac{64}{\Omega^2\: (u_+ - u_+^S)^4} \:. \label{Vppq}
\end{eqnarray}
In order to estimate the factor~$(u_+ - u_+^S)$ from below,
we use~(\ref{V1}) in the defining equation for~$u_+^S$, (\ref{usdef}),
\[ \frac{\Omega^2}{8}\: (u_+ - u_+^S)^4 \;\leq\;
|V(u_+^S)|\: (u_+ - u_+^S)^2 \;=\; \kappa^2 \:. \]
Using this inequality in~(\ref{Vpq}, \ref{Vppq}) and choosing~$\kappa$
sufficiently large, we obtain~(\ref{Vcd1},
\ref{Vcd2}) for all~$\tau \in (\frac{5 \pi}{16}, u_+^S]$.

On the interval~$[u_0, u_1]$, a short calculation using~(\ref{Vdef},
\ref{uJ01def}, \ref{range}) shows that
\begin{equation} \label{VpVpp2}
|V| \;\geq\; \frac{\lambda}{2}\:, \qquad
|V'|^2 \;\leq\; \frac{\lambda^3}{\kappa^6} \:,\qquad
|V''| \;\leq\; \frac{2 \lambda^2}{\kappa^4} \spc
{\mbox{on $[u_0, u_1]$}}\:,
\end{equation}
again proving~(\ref{Vcd1}, \ref{Vcd2}).

On the remaining interval~$(u_1, \frac{5 \pi}{16})$, we know
from the monotonicity of the potential and~(\ref{V1}) that
\[ |V(\tau)| \;\geq\;
\frac{\Omega^2}{8}\: \left(u_+-\frac{5 \pi}{16} \right)
\left(\pi-u_+-\frac{5 \pi}{16} \right)
\;\geq\; \frac{\pi^2}{2048}\: \Omega^2\:. \]
Furthermore, a short calculation using~(\ref{Vdef}) shows that
on~$(u_1, \frac{5 \pi}{16})$
\begin{equation} \label{VpVpp}
|V'| + |V''| \;\leq\; 4 \Omega^2 \spc {\mbox{on~$[u_1, \frac{\pi}{2}]$}}\:.
\end{equation}
We conclude that, choosing $\Omega_0$ sufficiently large,
we can again arrange that~(\ref{Vcd1}, \ref{Vcd2}) holds. \\[.8em]
{\bf{Second case:}} $k=0$ and $u_+ < \frac{3 \pi}{8}$. \\
On the interval~$[u_0, u_1]$, we can again use the estimate~(\ref{VpVpp2}).
Conversely, on the interval~$[u_1,\frac{3 \pi}{8}]$,
a short calculation shows that~$V''$ can be
bounded in terms of higher powers of the first derivatives; more precisely,
\[ |V''| \;\leq\; 10\:|V'|^{\frac{4}{3}}\:.  \]
This inequality allows us to deduce~(\ref{Vcd2})
from~(\ref{Vcd1}). Hence it remains to prove the inequality~(\ref{Vcd1})
on the interval~$[u_1, u_+^S]$.

On the interval~$[u_1, u_+^S]$, the potential~$V$ is either convex or else
at least the second derivative of~$V$ is large compared to~$|V'|^\frac{4}{3}$.
More precisely, by choosing~$\Omega_0$ sufficiently large, we can
arrange that
\begin{equation} \label{Vpp2}
V'' \;\geq\; -\kappa^{-\frac{2}{3}}\: V'^{\frac{4}{3}} \:.
\end{equation}
We shall derive an upper bound for~$\Delta u := u_+ - u_+^S$;
for ease in notation the subscript `$+$' will be omitted.
We rewrite~(\ref{Vpp2}) as
\[ \frac{d}{du} \left( V'^{-\frac{1}{3}} \right) \;=\;
-\frac{1}{3} \frac{V''}{V'^{\frac{4}{3}}} \;\leq\;
\frac{\kappa^{-\frac{2}{3}}}{3}\:. \]
We integrate from~$u^S$ to~$\tau \in [u^S, u]$
to obtain
\[ V'(\tau) \;\geq\; \left( V'(u^S)^{-\frac{1}{3}} +
\frac{\kappa^{-\frac{2}{3}}}{3}\: (\tau - u^S) \right)^{-3}\:. \]
Integrating~$\tau$ over the interval~$[u^S, u]$, we obtain for
$\Delta V := V(u)-V(u^S)$ the estimate
\[ \Delta V \;\geq\; \frac{3 \kappa^{\frac{2}{3}}}{2}\: V'(u^S)^\frac{2}{3}
\left( 1 - \frac{1}{(1+\alpha)^2} \right)
\qquad {\mbox{with}} \qquad
\alpha \;:=\; \frac{\kappa^{-\frac{2}{3}}}{3}\:
V'(u^S)^{\frac{1}{3}} \;\Delta u \:. \]
The inequality
\[ 1 - \frac{1}{(1+\alpha)^2} \;\geq\;
\frac{\alpha}{1+\alpha} \:, \]
gives
\[ \Delta V \;\geq\; \frac{V'(u^S)}{2}\:
\frac{\Delta u}{1+\alpha}\:. \]
By definition of~$u^S$, (\ref{usdef}), we know that~$\Delta V \!\cdot\!
(\Delta u)^2 = \kappa^2$. Hence, multiplying the last
inequality by~$(\Delta u)^2$, we obtain
\[ \kappa^2 \;\geq\; \frac{V'(u^S)}{2}\:
\frac{(\Delta u)^3}{1+\alpha}\:. \]
Using the definition of~$\alpha$ gives the inequality
\[ (\Delta u)^3 - \frac{2}{3} \:
\left(\kappa^{\frac{2}{3}}\: V'(u^S)^{-\frac{1}{3}}\right)^2 \Delta u
- 2 \left(\kappa^{\frac{2}{3}}\: V'(u^S)^{-\frac{1}{3}}\right)^3 \;\leq\; 0 \:. \]
Since the polynomial~$x^3 - \frac{2x}{3} - 2$ is positive for $x \geq 2$,
we conclude that
\[ \Delta u \;\leq\; 2 \:\kappa^{\frac{2}{3}}\: V'(u^S)^{-\frac{1}{3}}\:. \]
Using again the relation~$\Delta V (\Delta u)^2 = \kappa^2$, we get
an upper bound for~$\Delta V$,
\begin{equation} \label{dV}
\Delta V \;\geq\; \frac{\kappa^{\frac{2}{3}}}{4}\: V'(u^S)^\frac{2}{3}\:.
\end{equation}
This proves the inequality~(\ref{Vcd1}) at~$u=u^S$.

Next we shall show that~(\ref{Vcd1}) holds on the whole
interval~$(u_1, u^S]$. To this end, we introduce on this interval
the function $f$ by
\[ f \;=\; V'^2 + \frac{4^3}{\kappa^2}\: V^3 \:. \]
We saw above that~$f(u^S)<0$; our goal is to
show that~$f \leq 0$ on~$(u_1, u^S]$. Let~$(v, u^S]$ with~$u_1 \leq v < u^S$
be the maximal interval on which~$f$ is negative. We apply~(\ref{Vpp2}) to
obtain
\begin{eqnarray*}
f'(v) &=& V' \left( 2 V'' + \frac{3\cdot 4^3}{\kappa^2}\: V^2 \right) \\
&\geq& V' \left( -2 \kappa^{-\frac{2}{3}} V'^{\frac{4}{3}} + \frac{3\cdot 4^3}{\kappa^2}\: V^2 \right)
\;\geq\; 10\: \kappa^{-\frac{2}{3}} V'^{\frac{7}{3}} \;>\; 0\:,
\end{eqnarray*}
where in the last line we used that~$f(v) \leq 0$. The last inequality
contradicts the maximality of the interval~$(v, u^S]$ unless~$v=u_1$.
This concludes the proof in the second case.\\[.8em]
{\bf{Third case:}} $k \neq 0$. \\
On the interval~$(u_-^S, u_0]$, the proof of the second case goes through
without changes if we consider the integral backwards and
set~$u_1=u_0 - 1/(4 \sqrt{\Omega})$.
On the remaining interval~$[u_0, u_+^S]$, we can use the proof of the
first case and the second case after setting~$u_1=u_0 + 1/(4 \sqrt{\Omega})$.
\QED

\begin{Lemma} \label{lemmapr2}
For sufficiently large~$\Lambda$ and~$\Omega_0$,
the parameter~$L$ (as defined by~(\ref{Leq})) is  on the interval~$C$ for
all~$\Omega$, $\lambda$ in the range~(\ref{range}) bounded by
\[ L \;\leq\; \frac{3}{\sqrt{\Omega_0}} \:. \]
\end{Lemma}
{\Proof} Similar to~(\ref{Vpp}, \ref{Vp1}), one easily sees that~$V'$
and~$V''$ satisfy on~$C$ the bound~$|V'| + |V''| \leq 4 \Omega^2$.
On the other hand, it is clear
from the definition of~$u^I$ that~$|V| \geq \Omega^\frac{3}{2}$ on~$C$.
This immediately gives the lemma.
\QED

\begin{Lemma} \label{lemmapr3}
For sufficiently large~$\Lambda$ and~$\Omega_0$, the potential
\[ B(u) \;:=\; V(u) + \frac{1}{4 u^2} \]
satisfies on the interval~$P$ the inequality
\beq \label{Bb}
(u^J)^2\: (1+ \log^2 u^J)^2\:\|B\|_\infty \;\leq\; \frac{1}{8}\:.
\eeq
\end{Lemma}
{\Proof} A short calculation shows that~$B$ is bounded from above by
\[ |B(u)| \;\leq\; \Omega^2\: u^2 + 2 \lambda\:. \]
From the definition of~$u^J$, (\ref{uJ01def}), it is clear that for
large~$\lambda$, $|\log u^J| \leq \log \lambda$, and thus
\begin{eqnarray*}
(u^J)^2\: (1+ \log^2 u^J)^2\:\|B\|_\infty
&\leq& \frac{1}{64 \:\lambda\: \log^4 \lambda}\: (1+\log^2 \lambda)^2\: \|B\|_\infty \\
&\leq& \frac{1}{32\:\lambda}\: \|B\|_\infty
\;\stackrel{(\ref{Bb})}{\leq}\; \frac{1}{32\:\lambda} \left( \frac{\Omega^2}{\lambda} + 2 \lambda \right)
\;\leq\; \frac{1}{8} \:,
\end{eqnarray*}
where in the last step we again used~(\ref{range}).
\QED

\begin{Lemma} \label{lemmapr4}
For every~$\delta>0$ and~$k \in \Z$ there are parameters
$\kappa, \Lambda, \Omega_0 >0$ such that
for all~$\Omega, \lambda$ in the range~(\ref{range}),
\[ |I_\pm| \;\leq\; \delta \:. \]
\end{Lemma}
{\Proof} We choose~$\Omega_0$ so large that~$u_0 < \frac{\delta}{4}$.
Then clearly~$|I_-| \leq \delta$. Furthermore,
it is readily verified that the potential is increasing on the interval~$K :=
[\frac{\delta}{4}, \frac{\pi}{2} - \frac{\delta}{4}]$ at the following rate,
\[ V(v) - V(u) \;\geq\; c \: (v-u)^2\: \Omega^2
\qquad {\mbox{for all $u,v \in K$, $v>u$}} \;, \]
where~$c$ is independent of~$\lambda$ and~$\Omega$.
This implies that
\[ \left| [u_+^S, u_+] \cap K \right|
\;\leq\; c^{-\frac{1}{4}} \:\sqrt{\frac{\kappa}{\Omega}} \:, \]
because otherwise~(\ref{usdef}) would be violated. Furthermore,
the condition~$V(u^I) \leq \Omega^{\frac{3}{2}}$ implies that
\[ \left| [u_+, u^I] \cap K \right| \;\leq\; c^{-\frac{1}{2}}\:
\Omega^{-\frac{1}{4}}\:. \]
We conclude that by increasing~$\Omega_0$, we can arrange that
$|I_+ \cap K| \leq \frac{\delta}{2}$ and thus
$|I_+| \leq \delta$.
\QED

\section{Spectral Estimates for the Selfadjoint Problem}
\label{sec7}
\setcounter{equation}{0}
In this section we shall prove Lemma~\ref{lemma1}. We begin by
reducing the problem to an estimate for~$y_\lambda$.
\begin{Lemma} \label{lemma71}
Assume that for any given $k \in \Z$ and $\varepsilon>0$, there are
constants $\Lambda, \Omega_0 > 0$ such that
\begin{equation} \label{upper}
\int_0^{\frac{\pi}{2}} {\mbox{\rm{Im}}}\, y_\lambda \;\leq\; \varepsilon
\end{equation}
for all~$\Omega$ and~$\lambda$ in the range~(\ref{range}).
Then Lemma~\ref{lemma1} holds.
\end{Lemma}
{\Proof} According to the asymptotics~(\ref{asymptotics}) it suffices to
consider~$\lambda$ in the range~$\lambda > \Lambda \Omega$ for sufficiently
large~$\Lambda$. Let us consider the quotient~${\mbox{Re}}(y)/{\mbox{Im}}(y)$
in~(\ref{ps1}). According to Theorem~\ref{lemmay1} and Theorem~\ref{lemmay2},
there is a constant~$c>0$ such that for all~$\Omega > \Omega_0$ and~$\lambda > \Lambda \Omega$,
\begin{equation} \label{ReIm}
\frac{{\mbox{Re}}\, y}{{\mbox{Im}}\, y} \;>\; -c \spc {\mbox{on~$[u_0, u_+]$}}.
\end{equation}
On the interval~$[u_+, \frac{\pi}{2}]$, $\rho^2$ is convex, and using the identity
\[ \frac{{\mbox{Re}}\, y}{{\mbox{Im}}\, y} \;=\;
\frac{\rho^2}{w}\:{\mbox{Re}}\, y \;=\; \frac{1}{2w}\: (\rho^2)' \]
one sees that~${\mbox{Re}}(y)/{\mbox{Im}}(y)$ is monotone increasing.
We conclude that the inequality~(\ref{ReIm}) also holds at~$u=\frac{\pi}{2}$,
and thus
\[ -\frac{\pi}{2} \;<\; -\arctan c \;<\; \arctan \!\left( \frac{{\mbox{Re}}\,
y(\frac{\pi}{2})} {{\mbox{Im}}\, y(\frac{\pi}{2})} \right) \;<\; \frac{\pi}{2} \:. \]

Using the last bounds in~(\ref{ps1}) one sees that for two neighboring
eigenvalues, the phases must differ at least by $\delta := \frac{\pi}{2}-\arctan c$,
\begin{equation} \label{phdiff}
\varphi_{n+1} - \varphi_n \Big|_0^{\frac{\pi}{2}} \;\geq\; \delta \:.
\end{equation}
From~(\ref{ps2}) one sees that this inequality is also true for the states of
odd parity (with~$\delta=\pi$). Applying~(\ref{5B}) and the mean value theorem,
we conclude that there is~$\lambda \in [\lambda^\pm_n, \lambda^\pm_{n+1}]$ such that
\[ (\lambda^\pm_{n+1} - \lambda^\pm_n) \int_0^{\frac{\pi}{2}} {\mbox{Im}}\, y_\lambda
\;\geq\; \delta \:. \]
Hence the upper bound~(\ref{upper}) gives the desired gap estimate.
\QED

We establish~(\ref{upper}) by deriving separate estimates in the regions
$S$, $I_\pm$, $C$ and near the pole.
\begin{Lemma} \label{lemmasemi}
For any given $k \in \Z$ and $\varepsilon>0$, there are
constants $\Lambda, \Omega_0 > 0$ such that on the interval~$S$,
\[ |y_\lambda| \;\leq\; \varepsilon\:. \]
\end{Lemma}
{\Proof}
Differentiating the initial condition~(\ref{initc}) gives
\[ y_\lambda(u_0) \;=\; \frac{i}{2 \sqrt{|V(u_0)|}} + \frac{V'(u_0)}{4 V^2(u_0)} \:. \]
This can be estimated using Lemma~\ref{lemmapr1},
\begin{equation} \label{initest}
|y_\lambda(u_0)| \;\leq\; \frac{1}{2 \sqrt{|V(u_0)|}} \left(1 + \frac{|V'(u_0)|}{|V(u_0)|^\frac{3}{2}}
\right) \;\leq\; \frac{1}{\sqrt{|V(u_0)|}} \:.
\end{equation}
For given~$u \in S$, we compute~$y_\lambda(u)$ via~(\ref{ylam2}). This gives
rise to the estimate
\begin{eqnarray}
|y_\lambda(u)| &\leq& \frac{\rho^2(u_0)}{\rho^2(u)} \left| y_\lambda(u_0) + \frac{1}{2 y(u_0)} \right|
+ \frac{1}{2\,|y(u)|} + \frac{1}{\rho^2(u)} \int_{u_0}^u
\frac{|V-y^2|}{2\, |y|^2}\: \rho^2 \nonumber \\
&\leq& \frac{2\, {\mbox{Im}}\, y(u)}{|V(u_0)|}
+ \frac{1}{2\,|y(u)|} + {\mbox{Im}}\, y(u) \int_{u_0}^u
\frac{|V-y^2|}{2\, |y|^2}\: \frac{1}{{\mbox{Im}}\, y(u)} \:, \label{74a}
\end{eqnarray}
where in the last step we used~(\ref{initest}, \ref{usdef},
\ref{5C}).
According to Lemma~\ref{lemmapr1}, we can apply Theorem~\ref{lemmay1}.
We thus obtain for any~$K \in (0,1)$ the estimates
\begin{eqnarray}
|y| &\leq& 22\:\sqrt{|V|} \label{74b} \\
\left|y-i \sqrt{|V|} \right| &\leq& \sqrt{|V|}\: (20 K + \sqrt{K}) \;\leq\; 21 \sqrt{K} \sqrt{|V|}
\label{74c} \\
|V-y^2| &=& \left|(y-i \sqrt{|V|})(y+i \sqrt{|V|}) \right| \nonumber \\
&\leq& 21 \sqrt{K} \sqrt{|V|} \; 23\:\sqrt{|V|}
\;\leq\; 500\:|V|\:\sqrt{K}\:, \label{74d}
\end{eqnarray}
and, after choosing~$K < 1/42$, the inequality~(\ref{74c}) shows that
\beq \label{74e}
{\mbox{Im}}\, y \;\geq\; \frac{\sqrt{|V|}}{2}\:.
\eeq
Using the inequalities~(\ref{74b}, \ref{74e}) in~(\ref{74a}) gives
\[ |y_\lambda(u)| \;\leq\; \frac{44\: \sqrt{|V(u)|}}{|V(u_0)|} \:+\:
\frac{1}{\sqrt{|V(u)|}} \:+\: 22\, \sqrt{|V(u)|} \int_{u_0}^u
\frac{|V-y^2|}{|y|^2}\: \frac{1}{\sqrt{|V|}} \:, \]
and, since $V$ is monotonous, it follows using~(\ref{74e}, \ref{74d}) that
\begin{eqnarray*}
| y_\lambda(u)| &\leq&
\frac{45}{\sqrt{|V(u)|}} + 22 \int_{u_0}^u \: \frac{|V-y^2|}{|y|^2} \\
&\leq&
\frac{45}{\sqrt{|V(u)|}} + 88 \cdot 500\, \sqrt{K}\: \frac{\pi}{2}\:.
\end{eqnarray*}
The last expression can be made arbitrarily small according to
Lemma~\ref{lemmapr1}.
\QED

\begin{Lemma} \label{lemmainter}
For any given $k \in \Z$ and $\varepsilon>0$, there are
constants $\Lambda, \Omega_0 > 0$ such that on the intervals~$I_\pm$,
\[ |y_\lambda| \;\leq\; \varepsilon \:. \]
\end{Lemma}
{\Proof} We only consider the interval~$I_+$; the proof for~$I_-$ is analogous.
For any~$v \in (u_+^S, u_+]$, we compute~$y_\lambda$ via~(\ref{ylam}) with~$u=u_+^S$,
\[ y_\lambda(v) \;=\; -\frac{z^2(u_+^S)\: y_\lambda(u_+^S)}{z^2(v)}
-\frac{1}{z^2(v)} \int_{u_+^S}^v z^2 \:. \]
According to the definition of~$u_+^S$, (\ref{usdef}), we can apply Theorem~\ref{lemmay2}.
This gives the estimate
\[ |y_\lambda(v)| \;\leq\; |y_\lambda(u_+^S)| \:\frac{{\mbox{Im}}\,y(v)}{{\mbox{Im}}\,y(u_+^S)}
+ {\mbox{Im}}\,y(v)\:\int_{u_+^S}^v \frac{1}{{\mbox{Im}}\, y}
\;\leq\; c_2^2\:|y_\lambda(u_+^S)| + c_2^2\:(v-u_+^S)\:. \]
This can be made arbitrarily small according to
Lemma~\ref{lemmasemi} and Lemma~\ref{lemmapr4}.

If $v \in (u_+, u^I]$,
the change of~$y_\lambda$ on the interval~$(u_+, v)$ can be estimated
similarly using Lemma~\ref{lemmaconvex1}.
\QED

\begin{Lemma}
For any given $k \in \Z$ and $\varepsilon>0$, there are
constants $\Lambda, \Omega_0 > 0$ such that on the interval~$C$,
\[ |y_\lambda| \;\leq\; \epsilon \:. \]
\end{Lemma}
{\Proof} We again compute~$y_\lambda$ via~(\ref{ylam}). This gives
for any~$v \in C$ the estimate
\[ |y_\lambda(v)| \;\leq\; \frac{\rho^2(u^I)}{\rho^2(v)}\:|y_\lambda(u^I)|
+\frac{1}{\rho^2(v)} \int_{u^I}^v \rho^2 \:. \]
The first summand can be made arbitrarily small according to
Lemma~\ref{lemmainter} and Lemma~\ref{lemmaconvex1}, whereas the
second summand can be handled with Lemma~\ref{lemmaconvex2}
and Lemma~\ref{lemmapr2}.
\QED

\begin{Lemma}
For any given $k \in \Z$ and $\varepsilon>0$, there are
constants $\Lambda, \Omega_0 > 0$ such that for all~$\lambda$
and~$\Omega$ in the range~(\ref{range}),
\[ \int_{P \cup J} |y_\lambda| \;\leq\; \epsilon \:. \]
\end{Lemma}
{\Proof} A short calculation using~(\ref{Vdef}, \ref{uJ01def})
shows that on the interval~$J$,
\[ \frac{\lambda}{8 \kappa^2} \;\leq\; |V(u_0)| \;\leq\; \frac{\lambda}{\kappa^2} \;,\qquad
8\:\lambda\: \log^4 \lambda \;\leq\; |V(u^J)| \;\leq\; 64\:\lambda\: \log^4 \lambda \;,\qquad
|J| \;\leq\; \frac{\kappa}{\sqrt{\lambda}} \:. \]
In particular, $|V(u_0)|\:|J|^2 \leq 1$, and so we can
apply Theorem~\ref{lemmay2} to obtain on~$J$ the estimates
\begin{equation} \label{yvbnd0}
|y| \;\leq\; 16 c_2\: \sqrt{\lambda}\: \log^2 \lambda \;,\spc
{\mbox{Im}}\, y \;\geq\; \frac{1}{16 c_2 \kappa^2}\;
\frac{\sqrt{\lambda}}{\log^2 \lambda}\:.
\end{equation}
These estimates allow us to bound~$y_\lambda$ on~$J$ again
using~(\ref{ylam}). Namely, for all~$v \in J$,
\[ |y_\lambda(v)| \;\leq\; |y_\lambda(u_0)| \:\frac{{\mbox{Im}}\,y(v)}{{\mbox{Im}}\,y(u_0)}
+ {\mbox{Im}}\,y(v)\:\int_v^{u_0} \frac{1}{{\mbox{Im}}\, y} \:. \]
Estimating the factor $|y_\lambda(u_0)|$ by~(\ref{initest}), we obtain
\begin{equation} \label{yvbnd}
|y_\lambda(v)| \;\leq\; c_3 \frac{\log^4 \lambda}{\sqrt{\lambda}}
\spc {\mbox{on~$J$}}
\end{equation}
with~$c_3= 256 c_2 \kappa^3$. By increasing~$\Lambda$ this can be made arbitrarily small.

On the interval~$P$, we apply Lemma~\ref{thmpole} with~$C=64 c_2 \kappa^2 \log^4 \lambda$.
This gives the estimate
\begin{equation} \label{yvbnd2}
\frac{1}{c_4\: \log^p \lambda} \: \frac{1}{v \:\log^2 v}
\;\leq\; {\mbox{Im}}\; y(v) \;\leq\; \frac{c_4\: \log^p \lambda}{v \:\log^2 v}
\end{equation}
with~$p=14$ and a constant~$c_4$ which is independent of~$\Lambda$ and~$\Omega$.
We again estimate~$y_\lambda$ using~(\ref{ylam}),
\[ |y_\lambda(v)| \;\leq\; |y_\lambda(u^J)| \:\frac{{\mbox{Im}}\,y(v)}{{\mbox{Im}}\,y(u^J)}
+ {\mbox{Im}}\,y(v)\:\int_v^{u^J} \frac{1}{{\mbox{Im}}\, y} \:. \]
Estimating~$y(u^J)$ and~$y_\lambda(u^J)$ by~(\ref{yvbnd0}, \ref{yvbnd}) and
using~(\ref{yvbnd2}) we get for all~$v \in P$,
\begin{eqnarray*}
|y_\lambda(v)| &\leq& c_5 \:\frac{\log^{p+6} \lambda}{\lambda}\;
\frac{1}{v\:\log^2 v} \:+\: c_5
\frac{\log^{2p} \lambda}{v \:\log^2 v} \int_v^{u^J}
\tau\:\log^2 \tau\: d\tau
\end{eqnarray*}
for a suitable constant~$c_5$. This expression is not bounded
as~$v \searrow 0$. But the pole is integrable, and the calculation
\begin{eqnarray*}
\int_0^{u^J} \frac{dv}{v\: \log^2 v} &=& -\frac{1}{\log u^J} \\
\int_0^{u^J} \frac{dv}{v\: \log^2 v}
\int_v^{u^J} \tau\:\log^2 \tau\: d\tau
&=& \int_0^{u^J} d\tau\: \tau\:\log^2 \tau
\int_0^\tau \frac{dv}{v\: \log^2 v} \\
&=& -\int_0^{u^J} \tau\:\log \tau\:d\tau
\;=\; \frac{1}{4}\: (u^J)^2 \:(1-2 \log u^J)
\end{eqnarray*}
shows that, by increasing~$\Lambda$, we can make the resulting
integrals over~$P$ arbitrarily small.
\hspace*{2cm} \QED
This completes the proof of Theorem~\ref{thm3}.

\section{Slightly Non-Selfadjoint Perturbations}
\label{sec8}
\setcounter{equation}{0}
It remains to prove Theorem~\ref{thm1}. In preparation, we split up the
spheroidal wave operator as
\[ {\cal{A}} \;=\; {\cal{A}}_0 + W \]
with
\begin{eqnarray*}
{\cal{A}}_0 &=& -\frac{d}{d \cos \vartheta}\: \sin^2 \vartheta\:
\frac{d}{d \cos \vartheta}
+\frac{1}{\sin^2 \vartheta}({\mbox{Re}}\,\Omega \sin^2 \vartheta + k )^{2} \\[1ex]
W &=& 2i\: {\mbox{Im}}\, \Omega\:({\mbox{Re}}\,\Omega \sin^2 \vartheta + k )
- ({\mbox{Im}}\, \Omega)^2 \sin^2 \vartheta\:.
\end{eqnarray*}
The symmetric operator~${\cal{A}}_0$ clearly satisfies the hypothesis
of Theorems~\ref{thm2} and~\ref{thm3}, whereas the complex potential~$W$
is uniformly bounded according to assumption~(\ref{ocond}),
\begin{equation} \label{8e1}
|W| \;\leq\; 2 (k+1) c + c^2 \;=:\; \frac{\rho}{2}\:.
\end{equation}
Our method is to treat~$W$ as a slightly non-selfadjoint perturbation as
introduced by Kato~\cite[V.4.5]{Kato}; see in particular~\cite[Theorem~4.15a]{Kato}.
Unfortunately, the latter theorem is not quite strong enough for our purpose.
For clarity, we here repeat the basic ideas of Kato and give a detailed
proof of our main theorem.\\[.5em]
{\em{Proof of Theorem~\ref{thm1}.}} Throughout the proof, we restrict all
operators either to~${\cal{H}}_+$ or~${\cal{H}}_-$. Applying
Theorems~\ref{thm2} and~\ref{thm3} to the operator~${\cal{A}}_0$
and~$\gamma = 8 \rho$, we obtain for the eigenvalues~$0 \leq \lambda_1 < \lambda_2
< \cdots$ of~${\cal{A}}_0$ the estimates
\[ \lambda_{n+1} - \lambda_n \;>\; \gamma \spc
{\mbox{for all~$n \geq N$ and $\Omega \in \R$}}. \]
For all~$\lambda \not \in \sigma({\cal{A}}_0)$, the resolvent~$R_\lambda^0
:= (\lambda-{\cal{A}}_0)^{-1}$ exists and satisfies the bound
\begin{equation} \label{8e2}
\| R_\lambda^0 \| \;\leq\; \frac{1}{{\mbox{dist}}(\lambda, \sigma({\cal{A}}_0))}\:.
\end{equation}
Since the spectrum of~${\cal{A}}_0$ is real, we have in particular
\[ \|R_\lambda^0\| \;\leq\; \frac{1}{|{\mbox{Im}}\, \lambda|}\:. \]

\begin{figure}[tb]
\begin{center}
\input{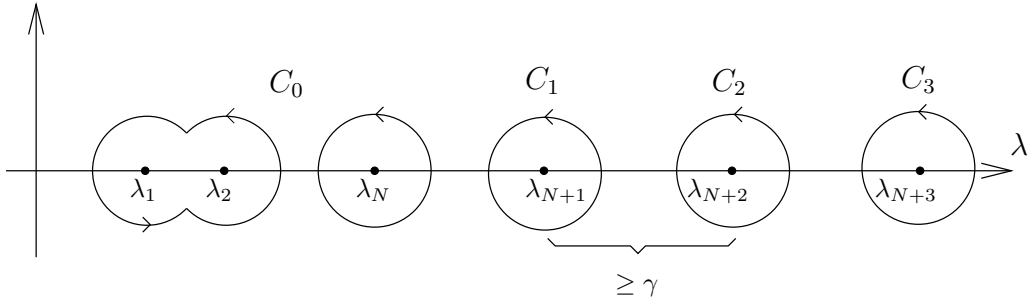}
\caption{The contours~$C_k$.}
\label{fig2}
\end{center}
\end{figure}

Around each~$\lambda_n$, we draw a circle of radius~$\rho$. The first~$N$ circles
may intersect, and we take the outermost lines to define the contour~$C_0$,
\[ C_0 \;=\; \partial (B_\rho(\lambda_1) \cup \ldots \cup B_\rho(\lambda_N) \:. \]
All the following circles do not intersect and give rise to the contours
\[ C_k \;=\; \partial B_\rho(\lambda_{N+k}) \;,\spc k \geq 1 \]
(see Figure~\ref{fig2}).
Since the distance of these contours to the spectral
points of~${\cal{A}}_0$ is at least~$\rho$, we have for~$\lambda$ on any of
these contours,
\[ \|R_\lambda^0 \:W \| \;\leq\; \|R_\lambda^0\| \: \| W \| \;\leq\;
\frac{1}{2}\:. \]
Hence the operator~$1+R_\lambda^0 W$ is invertible with a Neumann series.
We conclude that the resolvent~$R_\lambda := (\lambda-{\cal{A}})^{-1}
= ((\lambda-{\cal{A}}_0)\: (1-R_\lambda^0 W))^{-1} =
(1-R_\lambda^0 W)^{-1} R_\lambda^0$ exists and
\begin{equation} \label{8e3}
\|R_\lambda\| \;\leq\; 2\: \|R_\lambda^0\|\:.
\end{equation}
This allows us to introduce the operators~$Q_k$ as the following contour
integrals
\[ Q_k \;=\; \frac{1}{2 \pi i} \oint_{C_k} R_\lambda\: d\lambda\:. \]
The Cauchy integral formula together with the resolvent identity
\[ R_\lambda\: R_{\lambda'} \;=\; -\frac{1}{\lambda-\lambda'}\; (R_\lambda - R_{\lambda'}) \]
immediately yield that the operators~$Q_k$ are projectors onto invariant subspaces
of~${\cal{A}}$, and that they are holomorphic in~$\Omega$. Furthermore, they
are uniformly bounded because according to~(\ref{8e3}, \ref{8e2}) and the definition
of the contours,
\[ \|Q_k\| \;\leq\; \frac{1}{2 \pi} \oint_{C_k} 2\:\|R_\lambda^0\| \;\leq\; 2N\:. \]

\begin{figure}[tb]
\begin{center}
\input{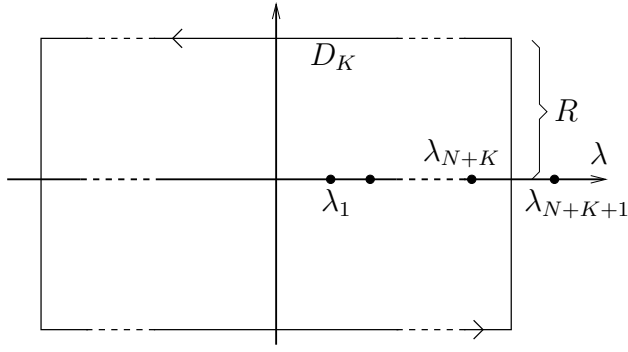}
\caption{The contours~$D_K$.}
\label{fig3}
\end{center}
\end{figure}

We introduce the operators~$P_K$ as the finite sums
\[ P_K \;=\; \sum_{k=0}^K Q_k \:. \]
For the unperturbed operator~${\cal{A}}_0$, we introduce similarly
the projectors~$Q_k^0$ and~$P_K^0$. Let us derive estimates for the
difference~$P_K - P_K^0$. We first write it as the contour integral
\beq \label{Rgl}
P_K - P_K^0 \;=\; \frac{1}{2 \pi i} \oint_{D_K} (R_\lambda - R_\lambda^0)\: d\lambda \:,
\eeq
where~$D_K$ is a rectangle with side lengths
$\lambda_{N+K} + \lambda_{N+K+1}$ and~$2R$ centered at the origin
(see Figure~\ref{fig3}).
Since~${\mbox{dist}}(D_k, \sigma({\cal{A}}_0)) > \rho$,
the inequality~(\ref{8e3}) again holds. Using the resolvent identity
\[ R_\lambda - R_\lambda^0 \;=\; R_\lambda\:W\: R_\lambda^0 \]
together with~(\ref{8e1}, \ref{8e2}), we obtain for any~$\lambda$ on the
contour~$D_K$ for sufficiently large~$R$ the estimate
\[ \|R_\lambda - R_\lambda^0\| \;\leq\; \frac{8 \rho}{\gamma^2 + 4\:
({\mbox{Im}}\,\lambda)^2}\:. \]
This inequality allows us to take in~(\ref{Rgl}) the limit~$R \to \infty$ to obtain
the estimate
\begin{equation} \label{8e4}
\|P_K - P_K^0\| \;\leq\; \frac{4 \rho}{\gamma} \;=\; \frac{1}{2}\:.
\end{equation}
This estimate can be improved if the operator~$P_K - P_K^0$ is restricted
to the range of~$P^0_L$, $L<K$. Namely, applying the bound
\[ \|R_\lambda^0 \:P_L^0 \| \;\leq\; \max_{n=1,\ldots, N+L} |\lambda-\lambda_n|^{-1} \]
to the resolvent identity gives for any~$\lambda$ on the contour~$D_K$
for sufficiently large~$R$ the bound
\[ \| (P_K - P_K^0)\:P_L^0 \| \;\leq\; \frac{8 \rho}{|\gamma + 2 i \:{\mbox{Im}}\, \lambda|
\; |\lambda-\lambda_{N+L}|}\:. \]
Substituting this estimate into the contour integral, taking the limit~$R \to
\infty$ and estimating the resulting integral as follows,
\[ \int_0^\infty \frac{dx}{\sqrt{a^2+x^2}\: \sqrt{b^2+x^2}}
\;\leq\; \int_0^{\sqrt{ab}} \frac{dx}{ab} \:+\:
\int_{\sqrt{ab}}^\infty \frac{dx}{x^2} \;\leq\; \frac{2}{\sqrt{ab}} \:, \]
we conclude that
\begin{equation} \label{8e5}
\|(P_K - P_K^0)\:P_L^0 \| \;\leq\; \frac{8 \rho}{\sqrt{\gamma\: (\lambda_{N+K} - \lambda_{N+L})}}
\;,\spc L<K.
\end{equation}

The inequality~(\ref{8e4}) allows us to determine the rank of the operators~$P_K$.
Namely, for every~$\Psi$ in the range of~$P_K^0$,
\[ \|P_K \Psi\| \;\geq\; \|P_K^0 \Psi\| - \|(P_K - P_K^0) \Psi \| \;\geq\;
\frac{1}{2}\: \|\Psi\|\:. \]
In particular, $\Psi$ is not in the kernel of~$P_K$. This shows that the rank
of~$P_K$ is greater or equal to the rank of~$P_K^0$. Interchanging the roles
of~$P_K$ and~$P_K^0$, we see that~$P_K$ and~$P_K^0$ have the same rank.
Since~$P_K^0$ is the projector on the eigenspaces of~${\cal{A}}_0$ corresponding
to the eigenvalues~$\lambda_1, \ldots, \lambda_{N+K}$, the dimension of its range is~$N+k$.
We conclude that~$Q_0$ is a projector on an $N$-dimensional invariant subspace
of~${\cal{A}}$ and $Q_1, Q_2, \ldots$ are projectors on $1$-dimensional
eigenspaces.

The inequalities~(\ref{8e4}, \ref{8e5}) imply completeness: Let~$\Psi \in
{\cal{H}}$ and~$\varepsilon>0$. Since the spectral projectors of the
unperturbed problem converge strongly (i.e.\ $s-\lim_{L \to \infty}P_L^0 = 1$),
there is~$L \in \N$ such that~$\|\Psi - P_M^0 \Psi \| < \varepsilon$
for all~$M \geq L$. According to~(\ref{8e4}), $\|P_K-\1\|
\leq \|P_K - P_K^0\| + \|P_K^0\| + \|1\| \leq 3$.
Hence for all~$K>L$,
\begin{eqnarray*}
\|(P_K - 1)\:\Psi \| &\leq& \|(P_K-1)\: (\Psi - P_L^0 \Psi)\|
\:+\: \|(P_K - P_K^0)\: P_L^0 \Psi\| \\
&\leq& 2 \varepsilon \:+\: \|(P_K - P_K^0)\: P_L^0 \Psi\| \:,
\end{eqnarray*}
and the estimate~(\ref{8e5}) shows that the last term can be made arbitrarily
small by choosing~$K$ sufficiently large.

It remains to prove the inequalities~(\ref{ange1}) and~(\ref{ang11}).
Combining the gap estimates of Theorem~\ref{thm3} with the fact
that~$\mu_k \in B_\rho(\lambda_{N+k})$, one immediately obtains~(\ref{ang11}).
The inequality~(\ref{ang11}) follows similarly from the
asymptotics~(\ref{asymptotics}).
\QED

\noindent
{\em{Acknowledgments:}}
We would like to thank Joel Smoller and Niky Kamran for helpful discussions,
and Johannes Kern for pointing out an error in an earlier version
of the proof of Lemma~\ref{lemmaconvex1}.
We are grateful to the ``Centre de Recherche
Math{\'e}mathiques,'' Montr{\'e}al, for support and hospitality. One of us (FF)
wants to thank the Max Planck Institute for Mathematics in the
Sciences, Leipzig, and the Morningside Center, Beijing, for hospitality and
support. Finally, we would like to thank the referee for helpful
comments.

\noindent
NWF I -- Mathematik, Universit{\"a}t Regensburg, 93040 Regensburg, Germany, \\
{\tt{Felix.Finster@mathematik.uni-regensburg.de}}, \\
{\tt{Harald.Schmid@mathematik.uni-regensburg.de}}

\end{document}